\documentclass[sigconf]{acmart}

\usepackage{booktabs} 
\usepackage{algorithm}
\usepackage[noend]{algpseudocode}
\usepackage{amsmath,amsthm,amssymb}
\usepackage{graphicx}
\usepackage{enumitem}
\usepackage{setspace}
\usepackage{color}
\usepackage[normalem]{ulem} 
\usepackage{xcolor}
\usepackage{subfigure} 
\theoremstyle{definition}

\usepackage{multirow}
\usepackage{subfigure} 

\usepackage{ textcomp }
\usepackage{ gensymb }
\usepackage{ wasysym }
\setcopyright{rightsretained}

\copyrightyear{2018}
\acmYear{2018}
\setcopyright{acmcopyright}
\acmConference[SIGIR '18]{The 41st International ACM SIGIR Conference on Research and Development in Information Retrieval}{July 8--12, 2018}{Ann Arbor, MI, USA}
\acmBooktitle{SIGIR '18: The 41st International ACM SIGIR Conference on Research and Development in Information Retrieval, July 8--12, 2018, Ann Arbor, MI, USA}
\acmPrice{15.00}
\acmDOI{10.1145/3209978.3210037}
\acmISBN{978-1-4503-5657-2/18/07}

 \newcommand{\squishlist}{
	\begin{list}{$\bullet$}
		{ \setlength{\itemsep}{0pt}
			\setlength{\parsep}{3pt}
			\setlength{\topsep}{3pt}
			\setlength{\partopsep}{0pt}
			\setlength{\leftmargin}{1.5em}
			\setlength{\labelwidth}{1em}
			\setlength{\labelsep}{0.5em} } }
	
	\newcommand{\squishlisttwo}{
		\begin{list}{$\bullet$}
			{ \setlength{\itemsep}{0pt}
				\setlength{\parsep}{0pt}
				\setlength{\topsep}{0pt}
				\setlength{\partopsep}{0pt}
				\setlength{\leftmargin}{2em}
				\setlength{\labelwidth}{1.5em}
				\setlength{\labelsep}{0.5em} } }
		
		\newcommand{\squishend}{
	\end{list}  }

\begin{document}
\title[The Rise of Guardians: Fact-checking URL Recommendation to Combat Fake News]{The Rise of Guardians: Fact-checking URL Recommendation \\ to Combat Fake News}


\author{Nguyen Vo and Kyumin Lee
}
\affiliation{
  \institution{Computer Science Department, Worcester Polytechnic Institute
  }
  \streetaddress{100 Road Institute
  }
  \city{Worcester, Massachusetts 01609, USA
  }
  \postcode{01609
  }
}
\email{{nkvo,kmlee}@wpi.edu}

%
%
%
%


\begin{abstract}
A large body of research work and efforts have been focused on detecting fake news and building online fact-check systems in order to debunk fake news as soon as possible. Despite the existence of these systems, fake news is still wildly shared by online users. It indicates that these systems may not be fully utilized. After detecting fake news, what is the next step to stop people from sharing it? How can we improve the utilization of these fact-check systems? To fill this gap, in this paper, we (i) collect and analyze online users called \emph{guardians}, who correct misinformation and fake news in online discussions by referring fact-checking URLs; and (ii) propose a novel fact-checking URL recommendation model to encourage the guardians to engage more in fact-checking activities. We found that the guardians usually took less than one day to reply to claims in online conversations and took another day to spread verified information to hundreds of millions of followers. Our proposed recommendation model outperformed four state-of-the-art models by 11\%$\sim$33\%. Our source code and dataset are available at \url{https://github.com/nguyenvo09/CombatingFakeNews}. 
\end{abstract}

%
%



\maketitle
\section{Introduction}
Fake news, misinformation, rumor or hoaxes are one of the most concerning problems due to their popularity and negative effects on society. Particularly, social networking sites (e.g., Twitter and Facebook) have become a medium to disseminate fake news. Therefore, companies and government agencies have paid attention to solving fake news. For example, Facebook has a plan to combat fake news\footnote{http://fortune.com/2017/10/05/facebook-test-more-info-button-fake-news/} and the FBI has investigated disinformation spread by Russia and other countries\footnote{http://bit.ly/FBIRussian}.

To verify correctness of information, researchers proposed to (i) employ experts, who can fact-check information \cite{zubiaga2016analysing}, (ii) use systems that can automatically check credibility of news \cite{shao2016hoaxy,liu2015real,hassan2017toward}; and build models to detect fake news \cite{castillo2011information, qazvinian2011rumor, kwon2013prominent, wu2015false, ma2015detect}. In 2016, Reporter Lab reported that the number of fact-checking websites went up by 50\%\footnote{http://reporterslab.org/global-fact-checking-up-50-percent}. However, fake news is still wildly disseminated on social media even when it has been debunked \cite{maddock2015characterizing, zhao2015enquiring, vosoughi2018spread}.

A recent report \cite{zignallab} showed that 86\% of American adults do not fact-check articles they read. A possible explanation for this is that people may trust content shared from their friends rather than other sources \cite{zignallab} or they may not have time to fact-check articles they read, or simply they may not know the existence of these fact-check websites. It means that merely debunking fake news is not enough, and these systems are not fully utilized. 

\begin{figure}[t]
	\centering
	\includegraphics[trim=2.1cm 6.4cm 10cm 20cm,clip,width=\linewidth,height=1.7in]{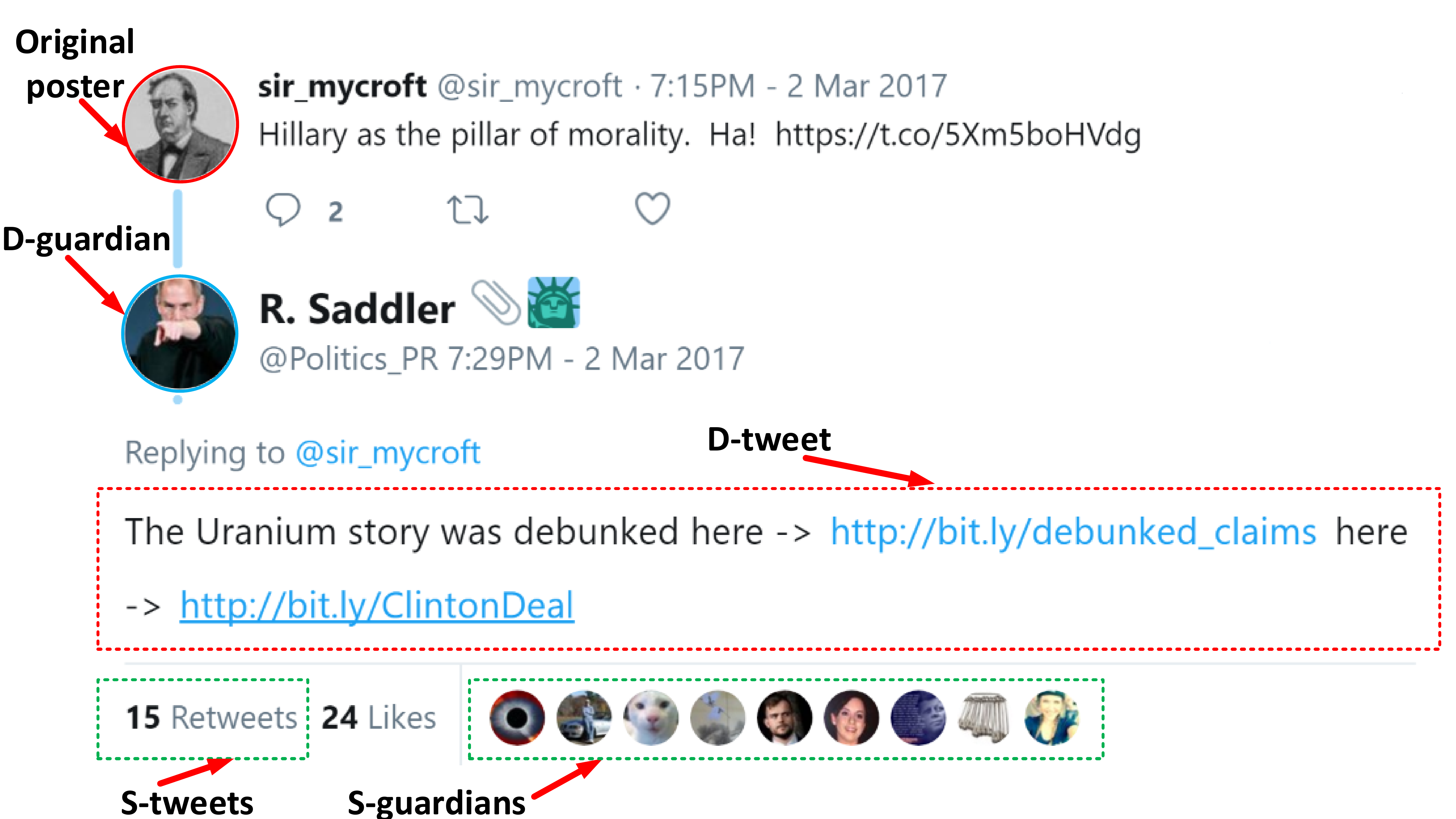}
	\caption{An example of fact-checking activity.}
	\label{fig:sample_fc_reply.PNG}
	\vspace{-10pt}
\end{figure}

Furthermore, it has been shown that once absorbing misinformation from fake news, individuals are less likely to change their beliefs even when the fake news are debunked. If the idea in the original fake news is especially similar to individuals' viewpoints, it will be even harder to change their minds \cite{ecker2010explicit, nyhan2010corrections}. Therefore, it is needed to deliver verified information quickly to online users before fake news reaches them. To achieve this aim, the volume of verified content should be large enough on social networks, so that online users may have a higher chance to be exposed to legitimate information before consuming fake news from other sources.

In this paper, we propose a framework to further utilize fact-checked content. Particularly, we collect a group of people and stimulate them to disseminate fact-checked content to other users. However, achieving the goal is challenging because we have to solve the two following problems: (\textbf{P1}) How can we find a group of people (e.g. online users) who are willing to spread verified news? (\textbf{P2}) How can we stimulate them to disseminate fact-checked news/information?

To deal with the first problem \textbf{(P1)}, we may deploy bots \cite{vo2017revealing,lee2011seven} to disseminate information but it may violate terms of services of online platforms due to abusing behavior. Another approach is to hire crowd workers \cite{lee2013crowdturfers} and cyber troops to shape public opinion \cite{bradshawtroops}. 
However, this approach may cost a lot of money and is difficult to deploy in larger scale due to monetary constraints. Inspired by \cite{hannak2014get}, we propose to rely on online users called \emph{guardians}, who show interests in correcting false claims and fake news in online discussions by embedding fact-checking URLs.
Figure \ref{fig:sample_fc_reply.PNG} illustrates who a guardian is and helps us to describe terminologies that we use in this paper. In the figure, two Twitter users have a conversation, in which a user @sir\_mycroft accused the Clinton foundation of accepting money from \textit{Uranium One} company in exchange for the approval of the deal between \textit{Uranium One} and Russian government in 2009. After just 15 minutes, this false accusation was debunked by a user @Politics\_PR, who referred to FactCheck.org and Snopes.com URLs as evidences to support his factual correction. We call such direct replies, which contain fact-checking URLs, \emph{direct fact-checking tweets} (\textbf{D-tweets}). Users, who posted D-tweets, are called \emph{direct guardians} (\textbf{D-guardians}). The user, to whom the D-guardian replied (i.e. @sir\_mycroft), is called an \textit{original poster}. In addition, we observed that @Politics\_PR's response was retweeted 15 times. We call these retweeters \emph{secondary guardians} (\textbf{S-guardians}), regardless of whether they added a comment or not inside the retweet. Their shares are called \emph{secondary tweets} (\textbf{S-tweets}). Both \textit{D-guardians} and \textit{S-guardians} are called \textit{guardians}, and both \textit{D-tweets} and \textit{S-tweets} are called \textit{fact-checking tweets}. In Section~\ref{sec:data_analysis}, we investigate whether both D-guardians and S-guardians play an important role in correcting claims and spreading fact-checked information. 

To cope with the second problem \textbf{(P2)}, we may directly ask the guardians to spread verified news like \cite{lee2014will}, but their response rate may be low because each guardian may be interested in different topics, and eventually, we may send unwanted requests to some of the guardians. Thus, we tackle the second problem by proposing a fact-checking URL recommendation model. By providing personalized recommendations, we may stimulate guardians' engagement in fact-checking activities toward spreading credible information to many other users and reducing the negative effects of fake news.

By addressing these two problems, we collect a large number of reliable guardians and propose a fact-checking URL recommendation model which exploits recent success in embedding techniques \cite{liang2016factorization} and utilizes auxiliary data to personalize fact-checking URLs for the guardians. Our main contributions are as follows:


\squishlist
\item We are the first work to utilize guardians, who can help spread credible information and recommend fact-checking URLs to the guardians as a pro-active way to combat fake news.
\item We thoroughly analyze who guardians are, their temporal behavior, and topical interests.
\item We propose a novel URL recommendation model, which exploits fact-checking URLs' content (i.e., linked fact-checking pages), social network structure, and recent tweets' content.
\item We evaluate our proposed model against four state-of-the-art recommendation algorithms. Experimental results show that our model outperforms the competing models by 11\%$\sim$33\%.
\squishend

\section{Related work}
In this section, we first summarize related work about fake news, rumors and misinformation. Then, we cover the prior work on URL recommendation on social network.

\subsection{Fake News, Rumors and Misinformation}
Although fake news on social media has been extensively studied, it still attracts the attention of communities due to its negative impact on society such as fake Russian Facebook ads and political events \cite{allcott2017social}. The majority of studies focused on classifying rumors to either true or false by exploiting different feature sets \cite{castillo2011information, qazvinian2011rumor, kwon2013prominent, wu2015false, ma2015detect} or by building deep learning models \cite{ruchansky2017csi, ma2016detecting}.
In natural disasters and emergency situations, misinformation was investigated as well \cite{kaigo2012social, gupta2013faking, zhao2015enquiring}. Several works attempted to detect rumors as soon as possible using disputed signals \cite{zhao2015enquiring, liu2015real}, leveraging network embedding \cite{wu2018tracing} and employing collective data sources \cite{popat2017truth, kim2018leveraging}. However, there is no work about combating fake news once it has been debunked.

Another direction is to detect or classify stances of users (e.g. supporting or denying) toward rumors \cite{qazvinian2011rumor,ferreira2016emergent} and to analyze how users' stances have changed over time 
 \cite{maddock2015characterizing, zubiaga2016analysing, li2016user}. In addition to studying rumors' content, researchers \cite{li2016user, kwon2013aspects} also analyzed who were involved in spreading those rumors. Since fake news can be viewed as misinformation, work about detecting content polluters \cite{lee2011seven}, social bots \cite{varol2017online} and malicious campaigns \cite{vo2017revealing} are also related to our work. The following two works \cite{hannak2014get, friggeri2014rumor} are perhaps the most closely related to our work. In particular, Hannak et al. \cite{hannak2014get} analyzed the social relationship between the fact-checking user and the fact-checked user in online conversations. \cite{friggeri2014rumor} employed fact-checking URLs in Snopes.com as a way to understand how rumors were spread on Facebook. Our work differs from the prior works \cite{hannak2014get, friggeri2014rumor} since we focus on guardians, their temporal behavior and topical interests, and propose a fact-checking URL recommendation model to personalize relevant fact-checking URLs. 

\subsection{URL Recommendation on Social Media}
Chen et al., \cite{chen2010short} proposed a content-based method to recommend URLs on Twitter.  \cite{abel2011analyzing} proposed hashtag-based, topic-based and entity-based methods to build user profiles for news personalization. By enriching user profiles with external data sources \cite{abel2013twitter, abel2011semantic}, Abel et al., improved URL recommendation results. Taking a similar content-based approach,  Yamaguchi et al., \cite{yamaguchi2013recommending} employed Twitter lists to recommend fresh URLs and \cite {guy2011personalized} tried to recommend URLs on streaming data. \cite{de2012chatter} proposed an SVM based approach to recommend URLs. Dong et al., \cite{dong2010time} exploited Twitter data to discover fresh websites for a web search engine. However, to the best of our knowledge there is no prior work employing matrix factorization models and auxiliary information to recommend URLs to guardians on Twitter. 
In addition to recommending URLs, researchers also focused on personalizing who to follow \cite{brzozowski2011should}, interesting tweets \cite{chen2012collaborative}, 
and Twitter lists \cite{rakesh2014personalized}.
\section{Data Collection}
\label{sec:data_collection}
In this section, we describe our data collection strategy. Unlike the prior work \cite{hannak2014get} which collected only a small number of D-tweets ($\sim$4000), we employed the Hoaxy system \cite{shao2016hoaxy} to collect a large number of both D-tweets and S-tweets. In particular, we collected 231,377 unique \textit{fact-checking tweets} from six well-known fact-checking websites - \textit{Snopes.com}, \textit{Politifact.com}, \textit{FactCheck.org}, \textit{OpenSecrets.org}, \textit{TruthOrfiction.com} and \textit{Hoax-slayer.net} -- via the APIs provided by the Hoaxy system which internally used Twitter streaming API. The collected data consisted of 161,981 D-tweets and 69,396 S-tweets (58,821 retweets of D-tweets and 10,575 quotes of D-tweets) generated from May 16, 2016 to July 7, 2017 ($\sim$ 1 year and 2 month). The number of our collected D-tweets is 40 times larger than the dataset used in the prior work \cite{hannak2014get}. 

Similar to the prior work, we removed tweets containing only base URLs  (e.g., snopes.com or politifact.com) or URLs simply pointing to the background information of the websites because the tweets containing these URLs may not reflect fact-checking enthusiasm and not contain fact-checking information. After filtering, we had 225,068 fact-checking tweets consisting of 157,482 D-tweets and 67,586 S-tweets posted by 70,900 D-guardians and 45,406 S-guardians. 7,167 users played both roles of D-guardians and S-guardians. The number of unique fact-checking URLs was 7,295. In addition, we also collected each guardian's recent 200 tweets. Table \ref{tbl:DataB} shows the statistics of our pre-processed dataset.

\begin{table}[t]
	\centering
	\resizebox{1.0\linewidth}{!}
	{
		\begin{tabular}{ccccc}
			\hline 
			$|$D-tweets$|$ & $|$S-tweets$|$ & $|$D-guardians$|$ & $|$S-guardians$|$ & $|$D\&S guardians$|$ \\ \hline
			157,482         & 67,586         & 70,900          & 45,406          & 7,167   \\ \hline  
		\end{tabular}
	}
	\caption{Statistics of our dataset.}
	\label{tbl:DataB}
	\vspace{-20pt}
\end{table}

\section{Characteristics of Guardians}
\label{sec:data_analysis}
From our dataset, we seek to answer the following research questions about guardians, their temporal behavior and topical interests. 


\begin{table}
	\centering
	\resizebox{1.0\linewidth}{!}
	{
	\begin{tabular}{|l|l|l|}
		\hline
		\multicolumn{3}{|c|}{Top15 D-guardians and \# of D-tweets}        \\ \hline
		RandoRodeo (450)              & stuartbirdman (318)          & \textcolor{red}{upayr} (214)                \\ \hline
		pjr\_cunningham (430)         & ilpiese (297)                & JohnOrJane (213)           \\ \hline
		TXDemocrat (384)              & BreastsR4babies (255)        & GreenPeaches2 (199)        \\ \hline
		\textcolor{red}{Jkj193741} (355)               & rankled2 (230)               & spencerthayer (195)        \\ \hline
		BookRageStuff (325)           & \_\_\_lor\_\_ (221)          & SaintHeartwing (174)       \\ \hline
		\multicolumn{3}{|c|}{Top 15 S-guaridans and \# of S-tweets} \\ \hline
		\textcolor{red}{Jkj193741} (294)               & MrDane1982 (49)              & LeChatNoire4 (35)          \\ \hline
		MudNHoney (229)               & pinch0salt (46)              & bjcrochet (34)             \\ \hline
		\_sirtainly (75)              & ActualFlatticus (42)         & \textcolor{red}{upayr} (33)                 \\ \hline
		Paul197 (66)                  & BeltwayPanda (36)            & 58isthenew40 (33)          \\ \hline
        Endoracrat (49)               & EJLandwehr (36)              & slasher48 (31)             \\ \hline
	\end{tabular}
	}
	\caption{Top 15 most active D-guardians and S-guardians, and associated \# of D-tweets and \# of S-tweets.}
	\label{tbl:highly-active-guardians}
	\vspace{-10pt}
\end{table}

%

\begin{table}[t]
	\centering
	\resizebox{1.0\linewidth}{!}
	{
		\begin{tabular}{|l|l|l|}
			\hline
			\multicolumn{3}{|c|}{Verified guardians and ($|$D-tweets$|$ vs. $|$S-tweets$|$)}               \\ \hline
			\textcolor{blue}{fawfulfan} (103-1)      & tomcoates (37-0)     & \textcolor{blue}{KimLaCapria} (27-3)    \\ \hline
			\textcolor{blue}{\textbf{OpenSecretsDC}} (37-30)  & \textcolor{blue}{\textbf{aravosis}} (29-8)      & PattyArquette (29-0)  \\ \hline
			\textcolor{blue}{\textbf{PolitiFact}} (41-17)     & \textcolor{blue}{\textbf{TalibKweli}} (27-8)    & NickFalacci (28-0)    \\ \hline
			\textcolor{blue}{\textbf{RobertMaguire}\_} (46-7) & rolandscahill (31-0) & AaronJFentress (28-0) \\ \hline
			\textcolor{blue}{jackschofield} (42-1)   & MichaelKors (30-0)   & \textcolor{blue}{ParkerMolloy} (26-1)   \\ \hline
		\end{tabular}
	}
	\caption{Top 15 verified guardians, and corresponding D-tweet and S-tweet count.}
	\label{tbl:high-active-verified-guardians}
	\vspace{-20pt}
\end{table}

\begin{figure}[t]
	\centering
	\subfigure[D-guardians' response time]{
		\label{fig:response_times_of_Dguardians}
		\includegraphics[trim=0 5 10 40,clip,width=0.3\linewidth,height=1.0in]{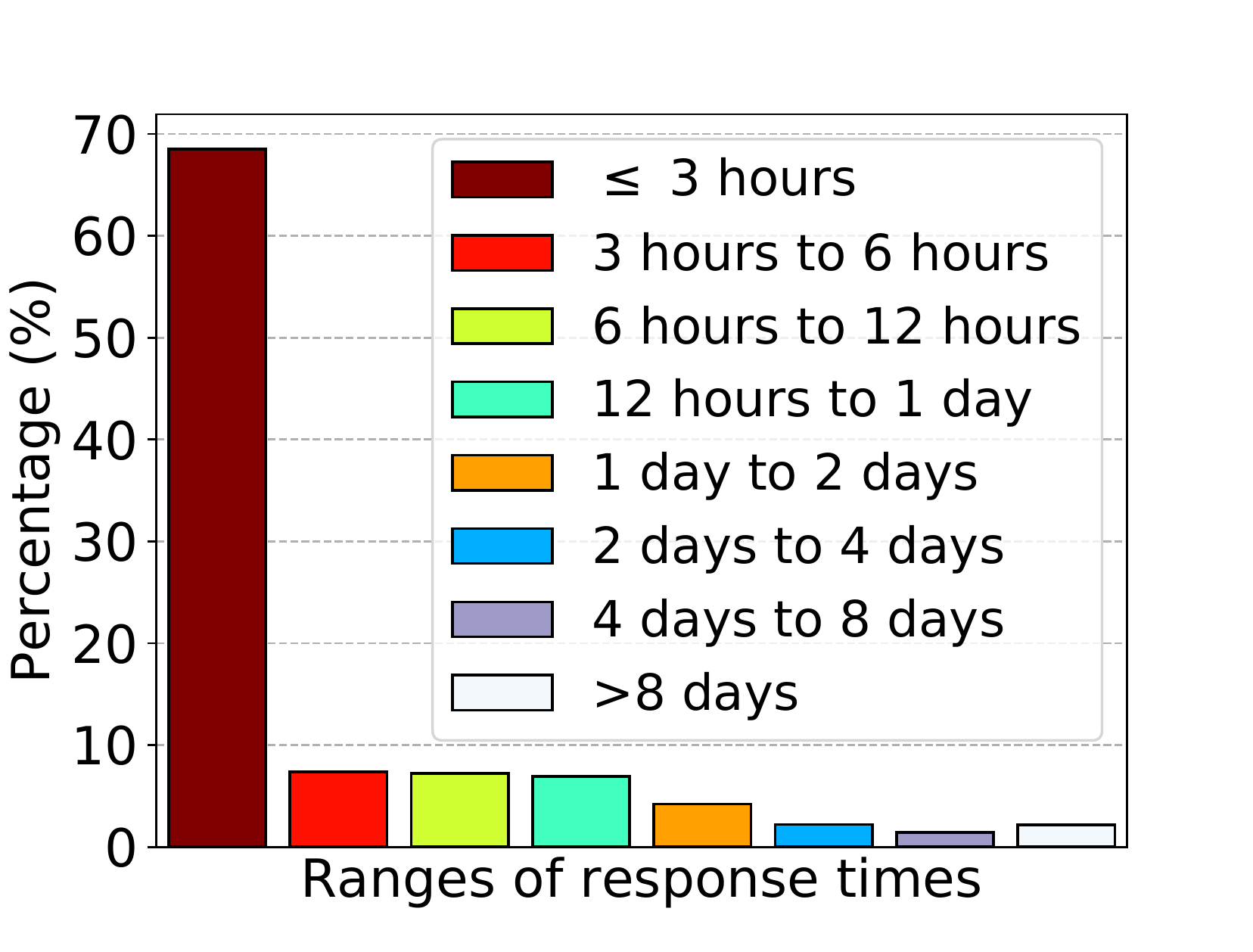}
	}
	\subfigure[S-guardians' response time]{
		\label{fig:response_times_S_guardians}
		\includegraphics[trim=0 5 10 45,clip,width=0.3\linewidth,height=1.0in]{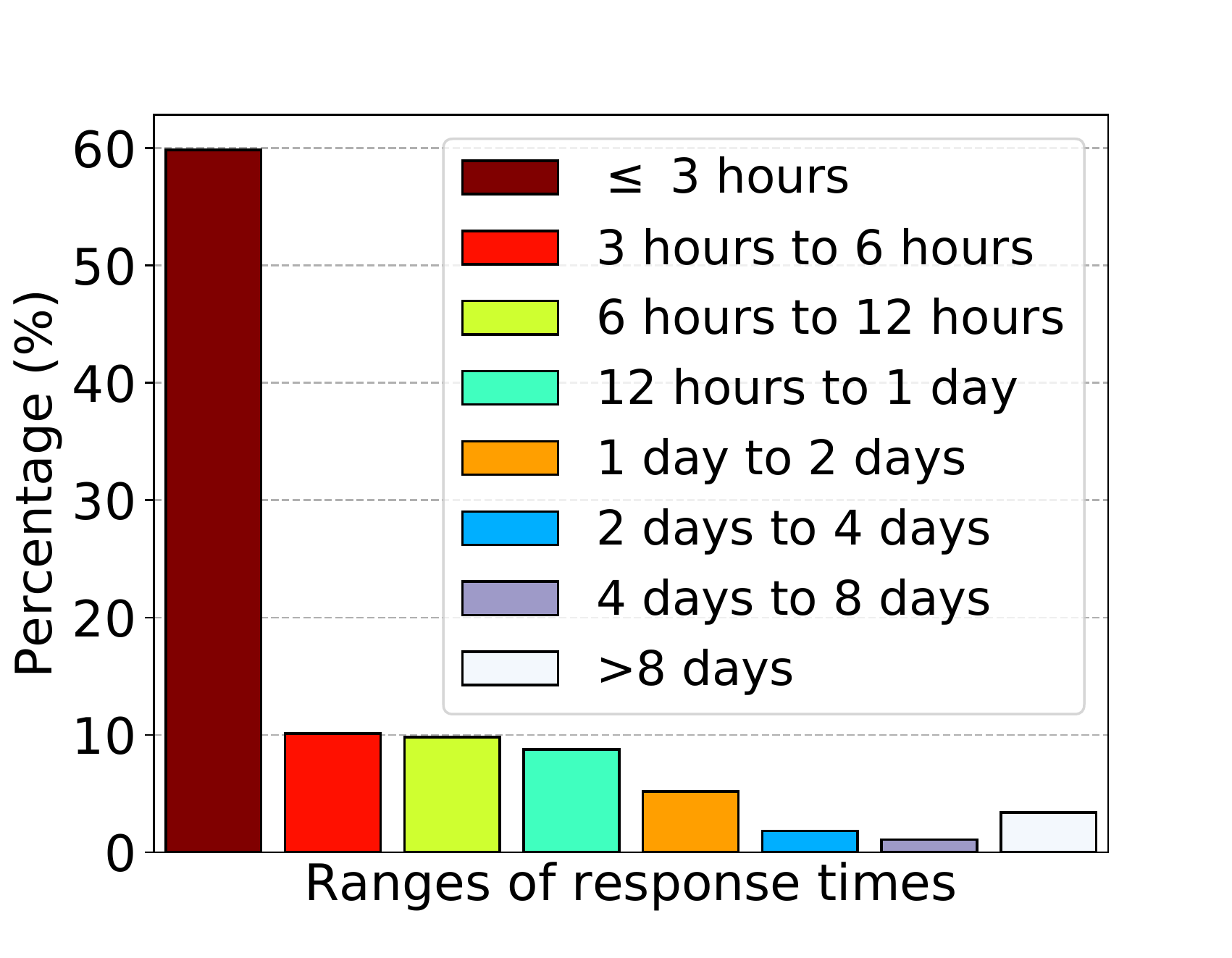}
	}
	\subfigure[S-tweets' inter-posting time]{
		\label{fig:how_fast_replies_spread}
		\includegraphics[trim=0 5 90 80,clip,width=0.3\linewidth,height=1.0in]{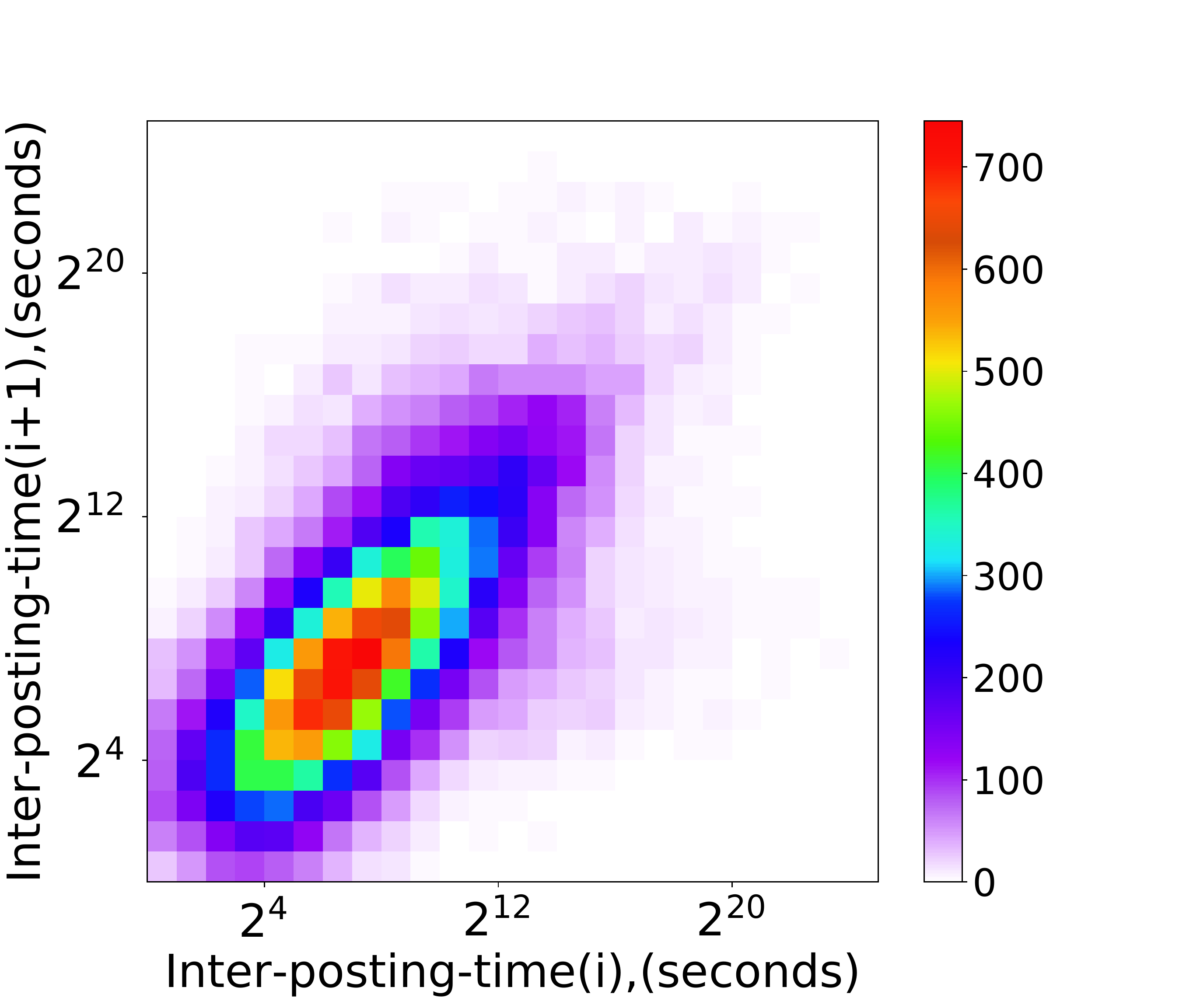}
	}
	\vspace{-5pt}
	\caption{Ranges of response time of D-guardians and S-guardians, and inter-posting time of S-tweets. The color in (c) indicates the number of pairs.}
	\vspace{-5pt}
\end{figure}

\smallskip
\noindent\textbf{Who are the guardians?}

\noindent As we have shown in the previous section, there were only 7,167 users (7\%) who behaved as both D-guardians and S-guardians, which indicates that guardians usually focused on either fact-checking claims in conversations (i.e., being D-guardians) or simply sharing credible information (i.e., being S-guardians). Since D-guardians and S-guardians played different roles, we seek to understand which group is more enthusiastic about its role. We created two lists - a list of the number of D-tweets posted by each D-guardian and a list of the number of S-tweets posted by each S-guardian --, excluding D\&S guardians who performed both roles. Then, by conducting One-sided MannWhitney U-test, we found that D-guardians were significantly more enthusiastic about their role than S-guardians (p-value<$10^{-6}$). We also found that even the D\&S guardians posted relatively larger number of D-tweets than S-tweets according to Wilcoxon one-sided test (p-value<$10^{-6}$).

The majority of guardians (85.3\%) posted only 1$\sim$2 fact-checking tweets. However, there were super active guardians, each of whom posted over 200 fact-checking tweets. Table \ref{tbl:highly-active-guardians} shows the top 15 most active D-guardians and S-guardians and the number of their D-tweets and S-tweets. As we can see, the most active D-guardians showed their strong enthusiasm for posting fact-checked content in online discussions.
Red-colored \emph{Jkj193741} and \emph{upayr} guardians were especially active in joining online conversations and spreading fact-checked information.

Next, we examined whether guardians have \emph{verified} Twitter accounts or are highly visible users, who have at least 5,000 followers. The verified accounts and highly visible users usually play an important role in social media since their fact-checking tweets can reach many audiences \cite{lee2014will, starbird2010pass}. Since the verified accounts are more trustworthy, their fact-checking tweets are often shared by many other users.
In our dataset, 2,401 guardians (2.2\%) had verified accounts. Table \ref{tbl:high-active-verified-guardians} shows the top 15 verified accounts. Interestingly, some of these verified accounts behaved as D\&S guardians, highlighted with the blue color in the table. Particularly, @PolitiFact, and @OpenSecretsDC, the official accounts of Politifact.com and OpenSecrets.org, frequently engaged in many online conversations. 
8,221 guardians (7.5\%) were highly visible users. Most top verified guardians, and many top S-guardians had a large number of followers. Altogether, S-tweets of the 45,406 S-guardians reached over 200 million followers.

Based on the analysis, we conclude that both D-guardians and S-guardians played important roles in terms of fact-checking claims and spreading the fact-checked news to the other users. Therefore, we need both types of guardians to spread credible information.

\begin{figure}[t]
	\centering
	\includegraphics[trim=50 8 120 70,clip,width=\linewidth,height=1.3in]{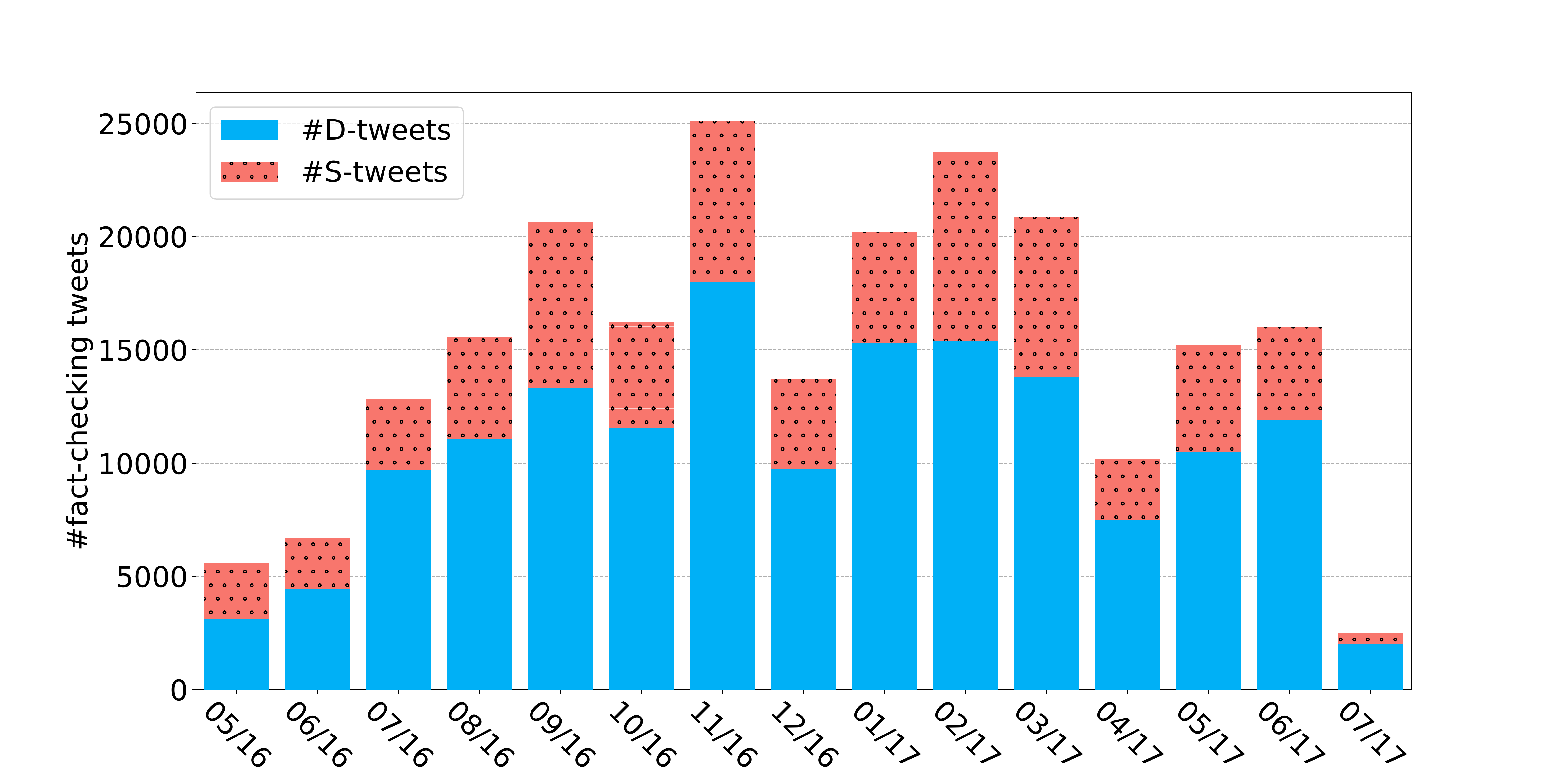}
	\caption{Temporal changes of \#fact-checking tweets}
	\label{fig:num_fact_checking_tweets_changes_over_time}
	\vspace{-5pt}
\end{figure}

\begin{figure}[t]
	\centering
	\includegraphics[trim=50 8 120 70,clip,width=\linewidth,height=1.5in]{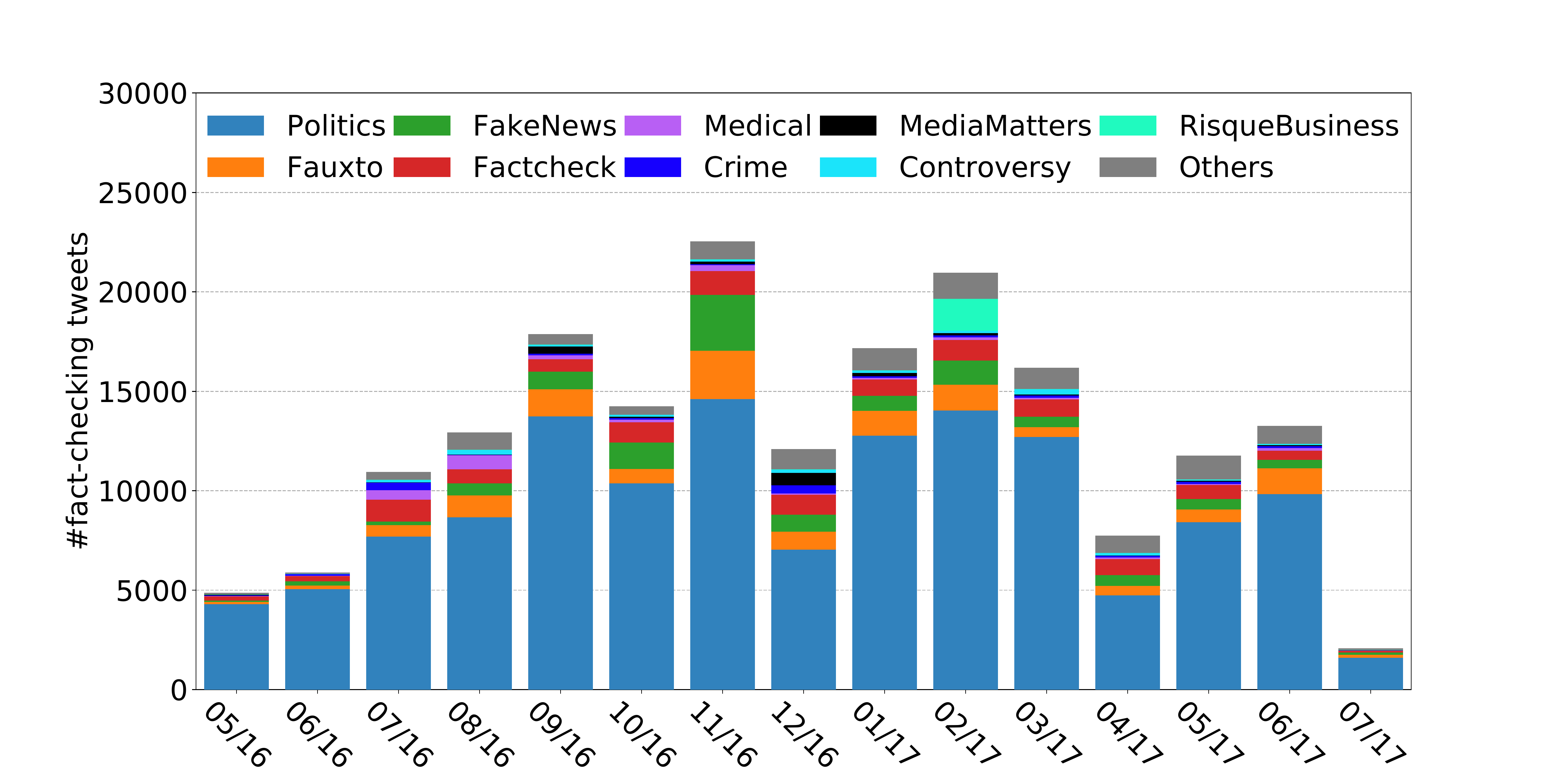}
	\caption{Topical changes of fact-checking tweets}
	\vspace{-10pt}
	\label{fig:topics_of_tweets_changes_over_time}
\end{figure}

\smallskip
\noindent\textbf{How quickly did guardians respond?}

\noindent\ To further understand activeness of guardians, we examined how quickly D-guardians posted their fact-checking URLs as responses to original posters' claims in online conversations. In particular, we measured response time of a D-tweet/D-guardian as a gap between an original poster's posting time and the fact-checking D-tweet's time. We collected all response time of {D-tweets}, grouped them and plotted a bar chart in Figure~\ref{fig:response_times_of_Dguardians}. The mean and median of response time were 2.26 days and 34 minutes, respectively. 90\% of D-tweets were posted within one day, indicating D-guardians quickly responded to the claims and expressed their enthusiasm by posting fact-checking URLs/tweets.

Similarly, we also measured response time of an S-tweet/S-guardian (Figure \ref{fig:response_times_S_guardians}) as a gap between D-tweet's posting time and the corresponding S-tweet's posting time. The mean and median of the response time were 3.1 days and 90 minutes, respectively. 88.5\% of S-tweets were posted within one day, indicating S-guardians also quickly responded and spread fact-checked information.


Finally, we measured S-guardians' inter-posting time to understand how long it took between two consecutive S-tweets, given the corresponding D-tweet. First, we grouped S-tweets based on each corresponding D-tweet, and sorted them in the ascending order of S-tweet creation time. Next, within each group, we computed inter-posting time $\delta_{i}$ as a gap between two consecutive S-tweets $i$ and $i+1$ and created pairs of inter-posting time $(\delta_{i}, \delta_{i+1})$. These pairs were merged across all the groups and were plotted in log2 scale in Figure \ref{fig:how_fast_replies_spread}. Overall, the average inter-posting time was 5 minutes, which means an S-tweet was posted once per 5 minutes by S-guardians after the corresponding D-tweet was posted. To sum up, both D-guardians and S-guardians were active and quickly responded to claims and fact-checked content.




\smallskip
\noindent\textbf{How did the volume of fact-checking tweets change over time? How did topics associated with fact-checking pages change over time?}

\noindent\ First, we examined the change in the number of fact-checking tweets (i.e., D-tweets and S-tweets) in each month between May 2016 and July 2017. Figure \ref{fig:num_fact_checking_tweets_changes_over_time} shows temporal changes of the number of fact-checking tweets. In the first 5 months, the number of fact-checking tweets increased gradually. In November 2016, the number of fact-checking tweets reached the peak (25,000 tweets) because of the US presidential election which happened on November 8, 2016. 
We also noticed that the number of D-tweets were larger than the number of S-tweets in every month which reflects that D-guardians were more active than S-guardians in online conversations (Wilcoxon one-side test p-value=$3.052\times 10^{-5}$). However, both D-guardians and S-guardians consistently posted and spread fact-checking tweets, respectively.

Next, we were interested in understanding what topics the fact-checking pages (linked by the URLs) were associated with and whether these topics changed over time. We first checked if a fact-checking website has categories, and if it did, we checked if we could automatically get the category information associated with each fact-checking page. For Snope pages, we identified each fact-checking page's topic by extracting the breadcrumb or tag information on the fact-checking page. We annotated PolitiFact pages' topic as \textit{politics} due to its political missions. In this analysis, we did not include fact-checking pages associated with the other four fact-checking websites because there were no explicit categories in content of the fact-checking pages, and their coverage was only 17.22\% (which would not contribute much to topical changes). Figure \ref{fig:topics_of_tweets_changes_over_time} shows temporal topical changes of fact-checking tweets in each month. Overall, \textit{politics} was the most popular in all months. Interestingly, fact-checking tweets under \textit{fauxtography}, \textit{fake news} and \textit{fact check} increased significantly in November 2016 (the month of US presidential election). In short, guardians' fact-checking activities were consistent over time, and their topical interests were mainly \textit{politics}, \textit{fauxtography} and \textit{fake news}.  
\begin{figure*}[h]
	\centering
	
	\subfigure[Fact-checking URLs]{
		\label{fig:top_fc_urls}
		\includegraphics[trim=0 20 50 40,clip,width=0.24\linewidth,height=1.4in]{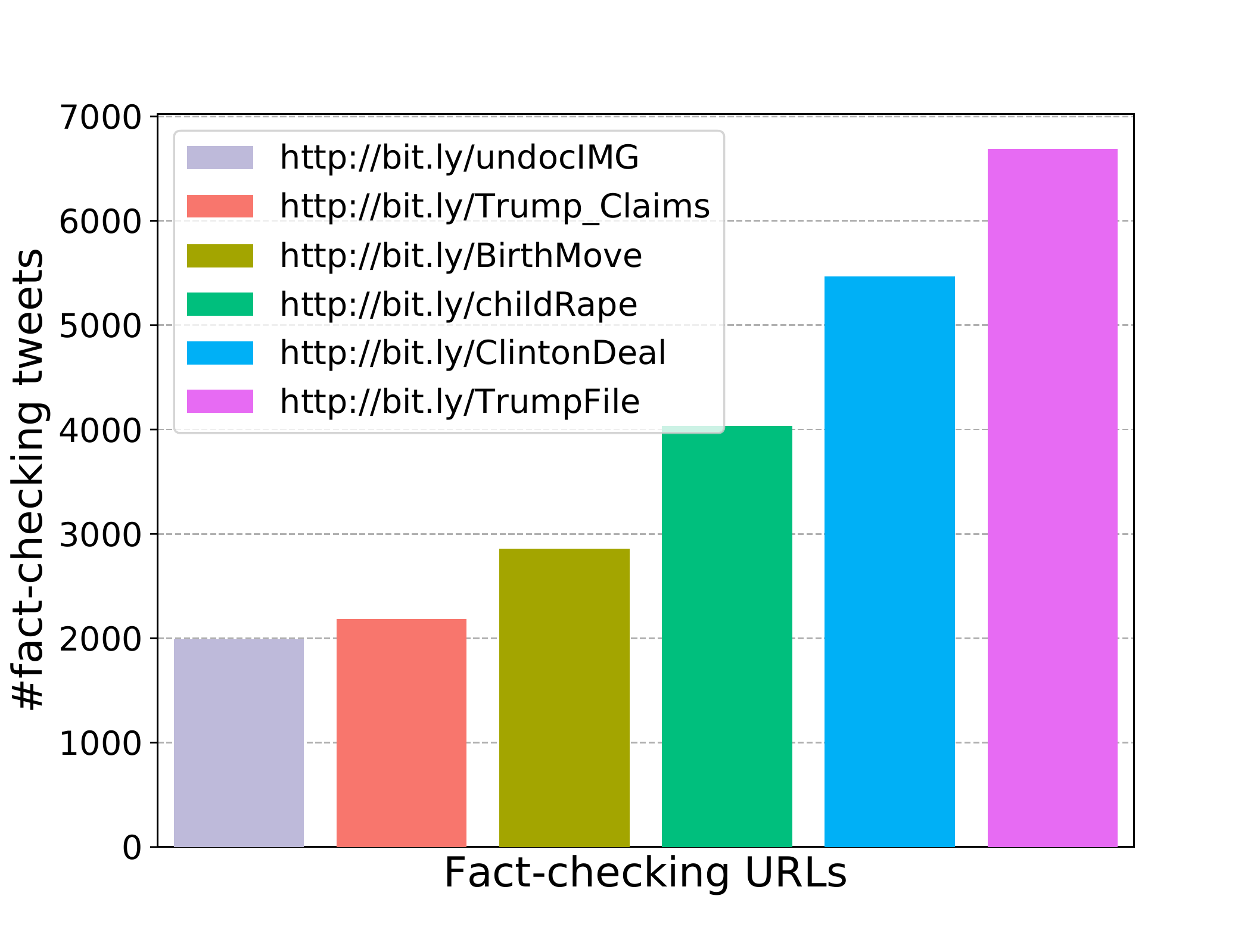}
	}
	\subfigure[Fact-checking websites]{
		\label{fig:URL-distribution-in-replies-tweets}
		\includegraphics[trim=90 80 80 100,clip,width=0.20\linewidth,height=1.4in]{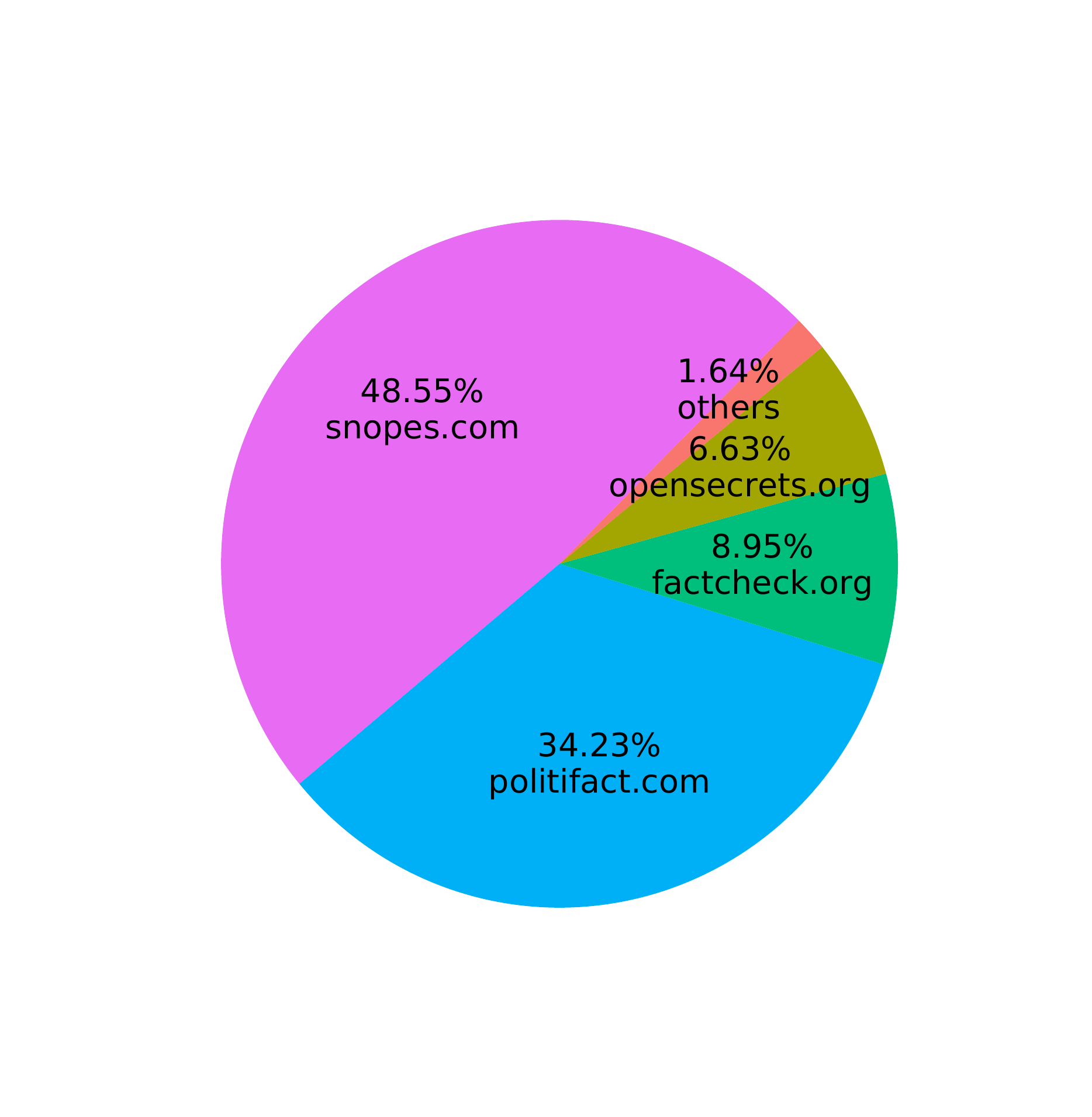}
	}
	\subfigure[Top words in fact-checking pages]{
		\label{fig:topics_in_urls}
		\includegraphics[trim=75 15 75 20,clip,width=0.22\linewidth, height=1.4in]{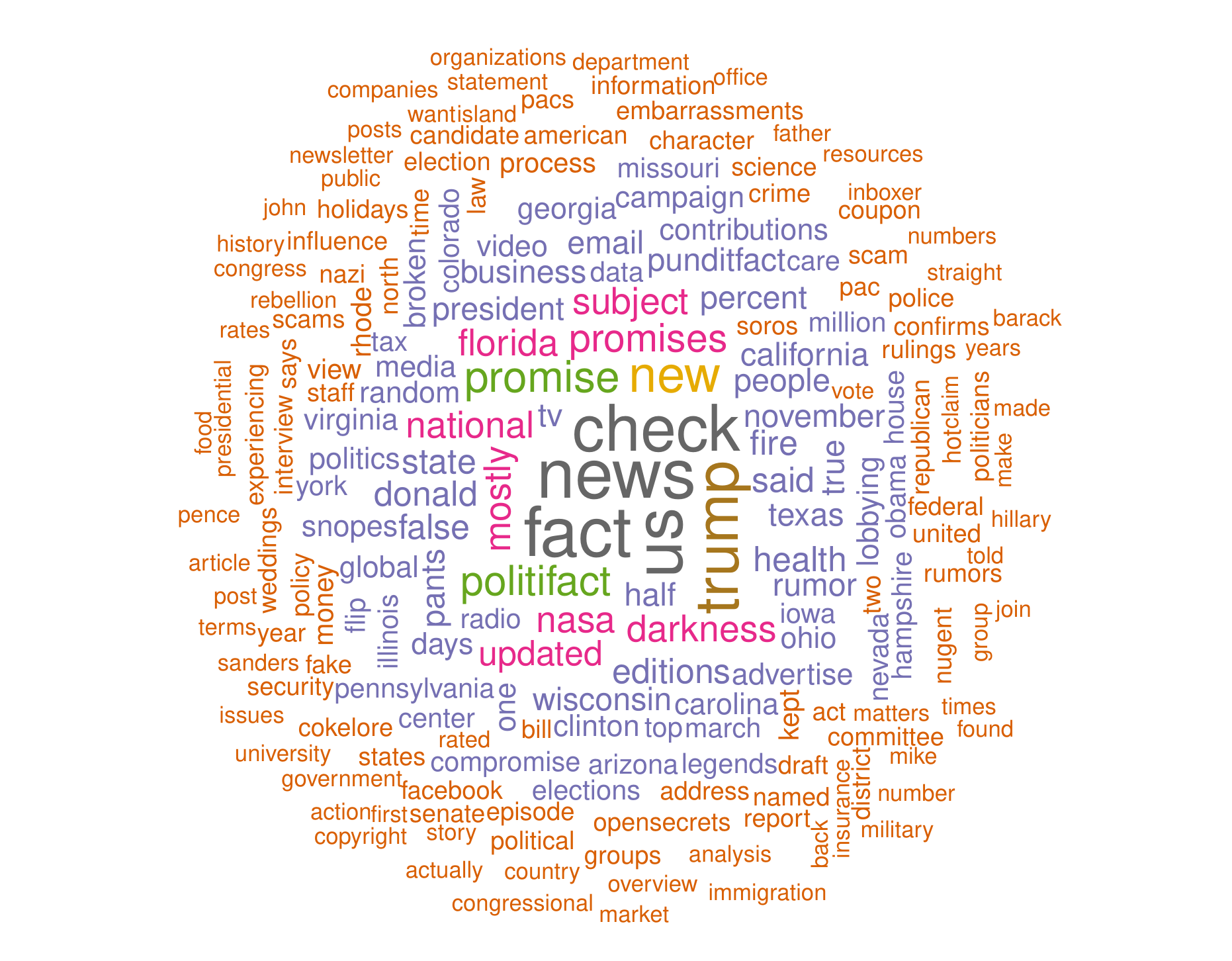}
	}
	\subfigure[Top words in recent 200 tweets]{
		\label{fig:topic_in_tweets}
		\includegraphics[trim=150 50 160 60,clip,width=0.22\linewidth, height=1.4in]{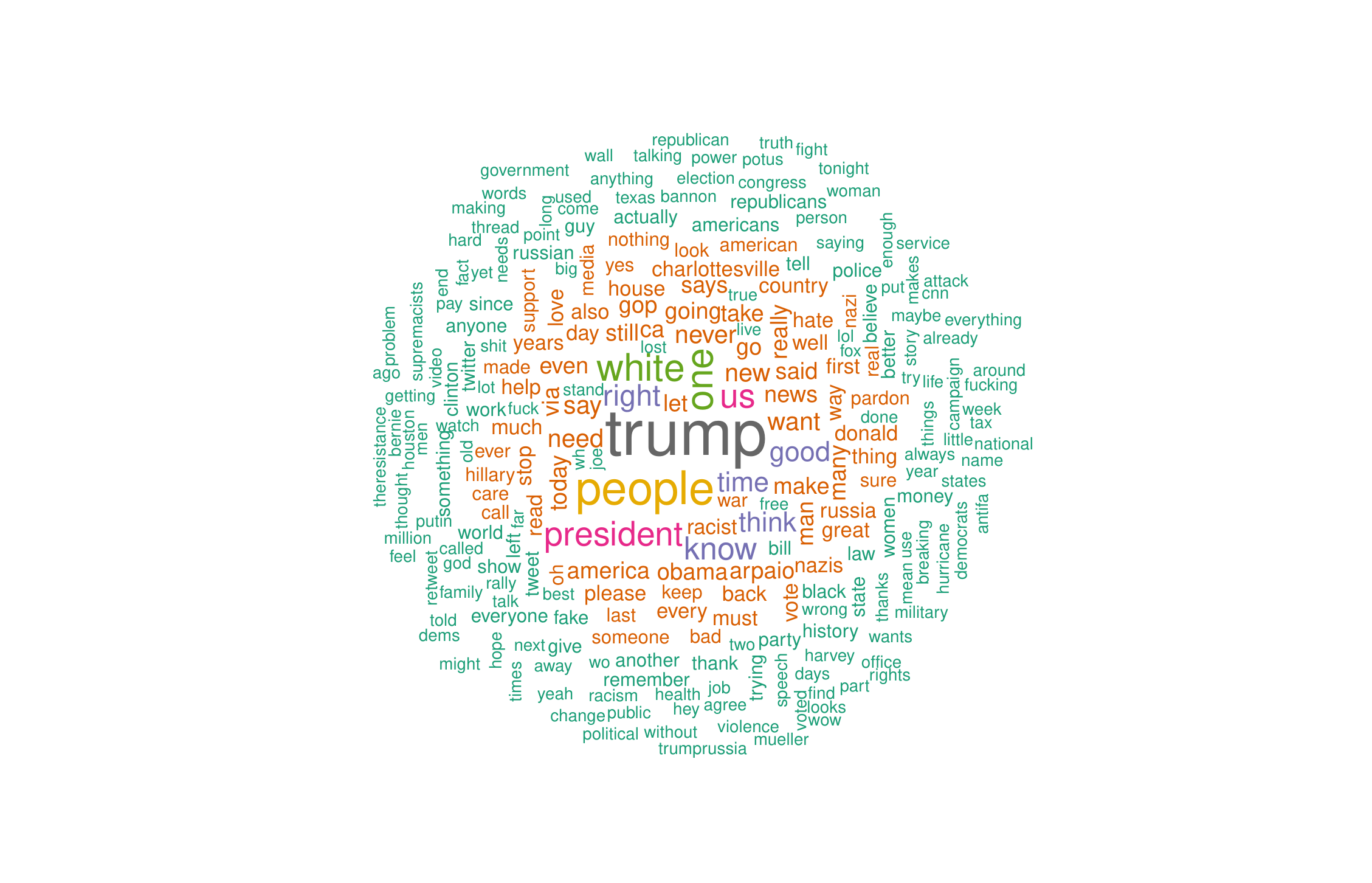}
	}
	\vspace{-10pt}
	
	\caption{(a) The most spread fact-checking URLs, (b) the most popular fact-checking websites, (c) the most important words in fact-checking pages linked by D-tweets and S-tweets and (d) the most important words in 200 recent tweets}
	\vspace{-5pt}
\end{figure*}

\smallskip
\noindent\textbf{What fact-checking URLs were spread most by the guardians? What fact-checking websites did guardians embed in fact-checking tweets? What were the most important terms used in the fact-checking pages and 200 recent tweets? }

\noindent Figure \ref{fig:top_fc_urls} shows the six most popular URLs embedded in fact-checking tweets. The URLs were related to Hillary Clinton and Donald Trump. Figure \ref{fig:URL-distribution-in-replies-tweets} shows what websites guardians used as references. Snopes.com was the most popular website, and politifact.com was the next frequently used one (48.55\% vs. 34.23\%). 

To answer the last question, given a fact-checking page linked by each of D-tweets and S-tweets, we extracted the main content after removing headers, footers and irrelevant content. Then, we selected the top 250 words according to tf-idf values. Similarly, given 200 recent tweets of each guardian, we first aggregated them to make a big document, removed non-English tweets, stop words, and URLs. Then, we selected the top 250 words according to tf-idf values. As shown in Figure \ref{fig:topics_in_urls} and \ref{fig:topic_in_tweets}, ``trump'' was mentioned often in both word clouds. Surprisingly, ``hillary'' and ``clinton'' were less frequently mentioned than Trump-related words. The figures also confirm that politics were one of popular topics, especially Trump-related news was one of popular claims.

%



\section{Fact-checking URL recommendation}
In the previous section, we found that the guardians are enthusiastic about credibility of information on social network and highly active in spreading fact-checked content. To encourage them to further engage in disseminating verified information, we propose a recommendation model to personalize fact-checking URLs. The aim of the recommendation model is to help guardians quickly access new interesting fact-checking URLs/pages so that they could embed them in their messages, correct unverified claims or misinformation, and spread fact-checked information. We use terms ``fact-checking URLs'' and ``URLs'', interchangeably.


\subsection{Problem Statement}
Let $\mathcal{N}=\{u_1, u_2, ..., u_N\}$ and $\mathcal{M}=\{\ell_1, \ell_2,...,\ell_M \}$ be a set of $N$ guardians and a set of $M$ fact-checking URLs, respectively. We view the action of embedding a fact-checking URL $\ell_j$ into a fact-checking tweet of guardian $u_i$ as an interaction pair $(u_i, \ell_j)$. We form a matrix $\textbf{X} \in \mathbb{R}^{N \times M}$ where $\textbf{X}_{ij}=1$ if the guardian $u_i$ posted a fact-checking URL $\ell_j$. Otherwise, $\textbf{X}_{ij}=0$. Our main goal is to learn a model that recommends similar URLs to guardians whose interests are similar. In particular, we aim to learn matrix $\textbf{U} \in \mathbb{R}^{N\times D}$, where each row vector $U_i^T \in \mathbb{R}^{D\times 1}$ is the latent representation of guardian $u_i$, and matrix $\textbf{V} \in \mathbb{R}^{D\times M}$, where each column vector $V_j \in \mathbb{R}^{D\times 1}$ is the latent representation of URL $\ell_j$. $D \ll min(M,N)$ is latent dimensions. Toward the goal, we propose our initial/basic matrix factorization model as follows:
\begin{equation}
\min_{\textbf{U,V}}  \| \Omega \odot (\textbf{X} - \textbf{U} \textbf{V}) \|_F^{2} + \lambda(\|\textbf{U} \|_F^2 + \| \textbf{V}\|_F^2)
\label{eq:basic}
\end{equation}

where $\Omega \in \mathbb{R}^{N \times M}$, and $\Omega_{ij}=1$ if $\textbf{X}_{ij}=1$. Otherwise, $\Omega_{ij}=0$. Operators $\odot$ and $\|.\|_F^2$ are Hadamard product and Frobenius norm, respectively. Finally, $\lambda$ is regularization factor to avoid overfitting. 
\subsection{Co-ocurrence model}
Now, we turn to extend our basic model in Eq.\ref{eq:basic} by further utilizing the interaction matrix $\textbf{X}$. Inspired by \cite{liang2016factorization, mikolov2013distributed}, we propose to regularize our basic model in Eq.\ref{eq:basic} by generating two additional matrices - URL-URL co-occurrence matrix and guardian-guardian co-occurrence matrix. Our main intuition of the extension is that a pair of URLs, which were posted by the same guardian, may be similar to each other. Likewise, a pair of guardians who posted the same URLs may be alike. To better understand our proposed models, we present the word embedding model as background information.

\subsubsection{Word embedding model}
Given a sequence of training words, word embedding models attempt to learn the distributed vector representation of each word. A typical example is \textit{word2vec} proposed by Mikolov et al. \cite{mikolov2013distributed}. Given a training word $w$, the main objective of the skip-gram model in \textit{word2vec} is to predict the \textit{context words} (i.e. the words that appear in a fixed-size context window) of $w$. Recently, it has been shown that training skip-gram model with negative sampling is similar to factorizing a word-context matrix named Shifted Positive Pointwise Mutual Information matrix ($SPPMI$) \cite{levy2014neural}. Given a word $i$ and its context word $j$, the value $SPPMI(i,j)$ is computed as follows:
\begin{equation}
	SPPMI(i,j) = max\{PMI(i,j) - log(s), 0\} \label{eq:sppmi}
\end{equation}
where $s\geq 1$ is the number of negative samples, and $PMI(i,j)$ is an element of Pointwise Mutual Information (PMI) matrix. $PMI(i,j)$ is estimated as $\log{\Big({{\#(i,j) \times |D|}\over {\#(i) \times \#(j)}}}\Big)$ where $\#(i,j)$ is the number of times that word $j$ appears in the context window of word $i$. $\#(i) = \sum_j \#(i,j)$, and $\#(j)=\sum_i\#(i,j)$. $|D|$ is the total number of pairs of word and context word. Note that $PMI(i,i)=0$ for every word $i$.

\subsubsection{URL-URL co-occurrence}
We generate a matrix $\textbf{R} \in \mathbb{R}^{M\times M}$ where $\textbf{R}_{ij}=SPPMI(\ell_i,\ell_j)$ based on co-occurrence of URLs. In particular, for each URL $\ell_i$ posted by a specific guardian, we define its context as all other URLs $\ell_j$ posted by the same guardian. Based on this definition, $\#(i,j)$ means the number of guardians that posted both URL $\ell_i$ and $\ell_j$. $\#(i,j)$ is also interpreted as the co-occurrence of URL $\ell_i$ and URL $\ell_j$. After that, we compute $PMI(\ell_i,\ell_j)$ and $SPPMI(\ell_i,\ell_j)$ based on Equation \ref{eq:sppmi} for all pairs of $\ell_i$ and $\ell_j$. 
\subsubsection{Guardian-Guardian co-occurrence}
Similarly, the context for each guardian $u_i$ is defined as all other guardians $u_j$ who posted the same URL with $u_i$. Then, $\#(i,j)$ is the number of URLs that both guardian $u_i$ and guardian $u_j$ commonly posted. Given this definition, we can generate a SPPMI matrix $\textbf{G} \in \mathbb{R}^{N\times N}$ where $\textbf{G}_{ij}=SPPMI(u_i,u_j)$. The same value of hyper-parameter $s$ is used for generating matrices $\textbf{R}$ and $\textbf{G}$.

\subsubsection{Regularizing matrix factorization with co-occurrence matrices} Our intuition is that URLs which are commonly posted by similar set of guardians are similar, and guardians who commonly posted the same set of URLs are close to each other. With that intuition, we propose loss function $\mathcal{L}_{XRG}$ -- a joint matrix factorization model of three matrices $\textbf{X}$, $\textbf{R}$ and $\textbf{G}$ as follows:
\begin{equation}
\begin{split}
\mathcal{L}_{XRG} &= \| \Omega \odot (\textbf{X} - \textbf{U} \textbf{V}) \|_F^{2} + \lambda(\|\textbf{U} \|_F^2 + \| \textbf{V}\|_F^2) \\ &+ \| \textbf{R}^{mask} \odot (\textbf{R} - \textbf{V}^T \textbf{K}) \|^2_F +  \| \textbf{G}^{mask} \odot (\textbf{G} - \textbf{U} \textbf{L}) \|^2_F
\end{split}
\label{eq:co-occurrence}
\end{equation}
where $\textbf{R}^{mask} \in \mathbb{R}^{M\times M}$, $\textbf{R}^{mask}_{ij}=1$ if $\textbf{R}_{ij}>0$. Otherwise, $\textbf{R}^{mask}_{ij}=0$. $\textbf{G}^{mask} \in \mathbb{R}^{N\times N}$, $\textbf{G}^{mask}_{ij}=1$ if $\textbf{G}_{ij}>0$. Otherwise, $\textbf{G}^{mask}_{ij}=0$. Two matrices $\textbf{K} \in \mathbb{R}^{D \times M}$ and $\textbf{L} \in \mathbb{R}^{D\times N}$ act as additional parameters. Although our work shares similar ideas with \cite{liang2016factorization}, there are three key differences between our model and \cite{liang2016factorization} as follows: (1) we omit bias matrices to reduce model complexity which is helpful in reducing overfitting, (2) additional matrix $\textbf{G}$ is factorized and (3) we do not regularize parameters $\textbf{K}$ and $\textbf{L}$.

\subsection{Integrating Auxiliary Information}
In addition, we propose auxiliary information which will be integrated with Eq.\ref{eq:co-occurrence} to improve URL recommendation performance. 

\subsubsection{Modeling social structure}
The social structure of guardians may reflect the homophily phenomenon indicating that guardians who follow each other may have similar interests in fact-checking URLs \cite{vo2017mrattractor}. To model this social structure of guardians, we first construct an unweighted undirected graph $G(V,E)$ where nodes are guardians, and an edge $(u_i,u_j)$ between guardians $u_i$ and $u_j$ are formed if  $u_i$ follows $u_j$ or $u_j$ follows $u_i$. In our dataset, in total, there were 1,033,704 edges in $G(V,E)$ (density=0.013898), which is 5.9 times higher than reported density in \cite{yang2012analyzing}, indicating dense connections between guardians. We represent $G(V,E)$ by using an adjacency matrix $\textbf{S} \in \mathbb{R}^{N\times N}$ where $\textbf{S}_{ij}=1$ if there is an edge $(u_i,u_j)$. Otherwise, $\textbf{S}_{ij}=0$. Second, we use Equation \ref{eq:social_structure} as a regularization term to make latent representations of connected guardians similar to each other.  Then, we formally minimize $\mathcal{L}_1$ as follows:
\begin{equation}
\mathcal{L}_1=\|\textbf{S}-\textbf{U}\textbf{U}^T\|_F^2
\label{eq:social_structure}
\end{equation}

\subsubsection{Modeling topical interests based on 200 recent tweets} In addition to social structure, the content of 200 recent tweets may reflect guardians' interests \cite{chen2010short, abel2011semantic,abel2011analyzing}. In Figure \ref{fig:topic_in_tweets}, 200 recent tweets of guardians contain many political words, which suggests us to enrich guardians' latent representation based on tweets' content.

For each guardian, we build a document by aggregating his/her 200 recent tweets and then employ the Doc2Vec model \cite{le2014distributed} to learn latent representations of the document. Doc2Vec is an unsupervised learning algorithm, which automatically learns high quality representation of documents. We use Gensim\footnote{https://radimrehurek.com/gensim/} as implementation of the Doc2Vec, set 300 as latent dimensions of documents, and train Doc2Vec model for 100 iterations. After training Doc2Vec model, we derive cosine similarity of every pair of learned vectors to create a symmetric matrix $\textbf{X}_{uu} \in \mathbb{R}^{N \times N}$, where $\textbf{X}_{uu}(i,j) \in [0;1]$ represents the similarity of document vectors of guardians $u_i$ and $u_j$. Intuitively, if two guardians have similar interests, their document vectors may be similar. Thus, we regularize guardians' latent representations to make them as close as possible by minimizing the following objective function:
\begin{equation}
\begin{split}
	\mathcal{L}_2&={1\over 2}\sum_{i=1,j=1}^{N}{\textbf{X}_{uu}(i,j)\| U_i^T - U_j^T \|^2} \\
	&=\sum_{i=1}^N U_i^T\textbf{D}_{uu}(i,i)U_i - \sum_{i=1,j=1}^N U_i^T\textbf{X}_{uu}(i,j)U_j \\
	&= Tr(\textbf{U}^T \textbf{D}_{uu} \textbf{U}) -  Tr(\textbf{U}^T \textbf{X}_{uu}\textbf{U}) = Tr(\textbf{U}^T \mathcal{L}_{uu} \textbf{U})
\end{split}
\end{equation}
where $\textbf{D}_{uu} \in \mathbb{R}^{N\times N}$ is a diagonal matrix with diagonal element $\textbf{D}_{uu}(i,i)=\sum_{j=1}^{N}\textbf{X}_{uu}(i,j)$. $Tr(.)$ is the trace of matrix, and $\mathcal{L}_{uu}=\textbf{D}_{uu}-\textbf{X}_{uu}$, which is a Laplacian matrix of the matrix $\textbf{X}_{uu}$.


\subsubsection{Modeling topical similarity of fact-checking pages}
We further exploit the content of fact-checking URLs (i.e., fact-checking pages) as an additional data source to improve recommendation quality. As we can see in Figure \ref{fig:topics_in_urls}, URLs' contents are mostly about politics. Intuitively, if the content of two URLs are similar (e.g. they are about Hillary Clinton's foundation as shown in Figure \ref{fig:sample_fc_reply.PNG}), their latent representations should be close. Exploiting the content of a fact-checking URL has been employed in \cite{abel2011semantic, wang2011collaborative}. In this paper, we apply a different approach, in which the Doc2Vec model is utilized to learn latent representation of URLs. Hyperparameters of the Doc2Vec model are the same as what we used for content of tweets. After training the Doc2Vec model, we derive the symmetric similarity matrix $\textbf{X}_{\ell\ell}\in \mathbb{R}^{M\times M}$ and minimize the loss function $\mathcal{L}_3$ in Equation \ref{eq:L3} as a way to regulate latent representation of URLs.
\begin{equation}
\begin{split}
\mathcal{L}_3&={1 \over 2}\sum_{i=1,j=1}^{M}{\textbf{X}_{\ell\ell}(i,j)\| V_i - V_j \|^2} \\
&=\sum_{i=1}^M V_i\textbf{D}_{\ell\ell}(i,i)V_i^T - \sum_{i=1,j=1}^M V_i\textbf{X}_{\ell\ell}(i,j)V_j^T \\
&=Tr(\textbf{V} (\textbf{D}_{\ell\ell} - \textbf{X}_{\ell\ell}) \textbf{V}^T) \\&= Tr(\textbf{V} \mathcal{L}_{\ell\ell} \textbf{V}^T)
\end{split}
\label{eq:L3}
\end{equation}
where $\textbf{D}_{\ell\ell} \in \mathbb{R}^{M\times M}$ is a diagonal matrix with elements on the diagonal $\textbf{D}_{\ell\ell}(i,i)=\sum_{j=1}^{M}\textbf{X}_{\ell\ell}(i,j)$ and $\mathcal{L}_{\ell\ell}=\textbf{D}_{\ell\ell}-\textbf{X}_{\ell\ell}$, which is the graph Laplacian of the matrix $\textbf{X}_{\ell\ell}$.

\subsection{Joint-learning fact-checking URL recommendation model}
Finally, we propose \textsc{GAU} - a joint model of \textbf{G}uardian-Guardian SPPMI matrix, \textbf{A}uxiliary information and \textbf{U}RL-URL SPPMI matrix. The objective function of our model, $\mathcal{L}_{GAU}$, is presented in Eq.\ref{eq:model12}:
\begin{equation}
	\begin{split}
	\min_{\textbf{U,V,L,K}}\mathcal{L}_{GAU} &= \| \Omega \odot (\textbf{X} - \textbf{U} \textbf{V}) \|_F^{2} + \lambda(\|\textbf{U} \|_F^2 + \| \textbf{V}\|_F^2)  \\ &+ \| \textbf{R}^{mask} \odot (\textbf{R} - \textbf{V}^T \textbf{K}) \|^2_F \\&+  \| \textbf{G}^{mask} \odot (\textbf{G} - \textbf{U} \textbf{L}) \|^2_F \\ &+ \alpha \times \|\textbf{S}-\textbf{U}\textbf{U}^T\|_F^2 \\ &+ \gamma\times Tr(\textbf{U}^T \mathcal{L}_{uu} \textbf{U}) \\&+ \beta\times Tr(\textbf{V} \mathcal{L}_{\ell\ell} \textbf{V}^T)
	\end{split}
	\label{eq:model12}
\end{equation}
where $\alpha, \gamma, \beta, \lambda$ and shifted negative sampling value $s$ are hyper parameters, tuned based on a validation set. We optimize $\mathcal{L}_{GAU}$ by using gradient descent to iteratively update parameters with fixed learning rate $\eta=0.001$. The details of the optimization algorithm are presented in Algorithm \ref{alg:xrg_solver}. After learning \textbf{U} and \textbf{V}, we estimate the guardian $u_i$'s preference for URL $\ell_j$ as: $\hat{r}_{i,j}\approx U_iV_j$. The final URLs recommended for a guardian $u_i$ is formed based on ranking:
\begin{equation}
u_i: \ell_{j_1} > \ell_{j_2} > ... >\ell_{j_M} \rightarrow \hat{r}_{i,j_1} > \hat{r}_{i,j_2} > ... > \hat{r}_{i,j_M}
\end{equation}
The derivatives of loss $\mathcal{L}_{GAU}$ with respect to parameters $\textbf{U}$, $\textbf{V}$, $\textbf{K}$ and $\textbf{L}$ are as follows:
\begin{equation}
\begin{split}
{\partial{\mathcal{L}_{GAU}} \over \partial{\textbf{U}} } = -2(\Omega \odot \Omega \odot (\textbf{X}-\textbf{UV}))\textbf{V}^T + 2\lambda\times (\textbf{U}) \\ -2(\textbf{G}^{mask} \odot \textbf{G}^{mask} \odot (\textbf{G}-\textbf{UL}))\textbf{L}^T \\ -2\alpha( (\textbf{S} - \textbf{U}\textbf{U}^T + (\textbf{S} - \textbf{U}\textbf{U}^T)^T)  \textbf{U}) \\ +\gamma \times (\mathcal{L}_{uu} + \mathcal{L}_{uu}^T)\textbf{U}\\
{\partial{\mathcal{L}_{GAU}} \over \partial{\textbf{V}} } = -2\textbf{U}^T(\Omega \odot \Omega \odot (\textbf{X}-\textbf{UV})) + 2\lambda\times ( \textbf{V}) \\ -2\textbf{K}(\textbf{R}^{mask} \odot \textbf{R}^{mask} \odot (\textbf{R}-\textbf{V}^T\textbf{K}))^T \\ +\beta \times \textbf{V}(\mathcal{L}_{\ell\ell} + \mathcal{L}_{\ell\ell}^T)\\
{\partial{\mathcal{L}_{GAU}} \over \partial{\textbf{L}} } = -2\textbf{U}^T(\textbf{G}^{mask} \odot \textbf{G}^{mask} \odot (\textbf{G}-\textbf{UL})) \\
{\partial{\mathcal{L}_{GAU}} \over \partial{\textbf{K}} } = -2\textbf{V}(\textbf{R}^{mask} \odot \textbf{R}^{mask} \odot (\textbf{R}-\textbf{V}^T\textbf{K}))
\end{split}
\label{eq:updating_rule_model12}
\end{equation}

\begin{algorithm}[H]
	\caption{\textsc{GAU Optimization algorithm}}
	\small
	\label{alg:xrg_solver}
	\begin{algorithmic}[1]
		\Statex \textbf{Input}: Guardian-URL interaction matrix $\textbf{X}$, URL-URL SPPMI matrix $\textbf{R}$, Guardian-Guardian SPPMI matrix $\textbf{G}$, social structure matrix $\textbf{S}$, Laplacian matrix $\mathcal{L}_{uu}$ of guardians, Laplician matrix $\mathcal{L}_{\ell\ell}$ of URLs, binary matrices $\Omega$, $\textbf{R}^{mask}$ and $\textbf{G}^{mask}$ as indication matrices.
		\Statex \textbf{Output}: $\textbf{U}$ and $\textbf{V}$
		\State Initialize $\textbf{U}$, $\textbf{V}$, $\textbf{K}$ and $\textbf{L}$ with Gaussian distribution $\mathcal{N}(0,0.01^2)$, $t \leftarrow 0$
		\While{Not Converged}
		\State Compute ${\partial{\mathcal{L}_{GAU}} \over \partial{\textbf{U}} }$, ${\partial{\mathcal{L}_{GAU}} \over \partial{\textbf{V}} }$, ${\partial{\mathcal{L}_{GAU}} \over \partial{\textbf{L}} }$ and ${\partial{\mathcal{L}_{GAU}} \over \partial{\textbf{K}} }$ in Eq.\ref{eq:updating_rule_model12}
		\State $\textbf{U}_{t+1} \leftarrow \textbf{U}_t - \eta{\partial{\mathcal{L}_{GAU}} \over \partial{\textbf{U}} } $
		\State $\textbf{V}_{t+1} \leftarrow \textbf{V}_t - \eta{\partial{\mathcal{L}_{GAU}} \over \partial{\textbf{V}} } $
		\State $\textbf{L}_{t+1} \leftarrow \textbf{L}_t - \eta{\partial{\mathcal{L}_{GAU}} \over \partial{\textbf{L}} } $
		\State $\textbf{K}_{t+1} \leftarrow \textbf{K}_t - \eta{\partial{\mathcal{L}_{GAU}} \over \partial{\textbf{K}} } $
		\State $t \leftarrow t+1$
		\EndWhile
		\Return \textbf{U} and \textbf{V}
	\end{algorithmic}
\end{algorithm}
\vspace{-10pt}

\section{Evaluation}
In this section, we thoroughly experiment our proposed $GAU$ model. In particular, we aim to answer the following research questions:
\squishlist
\item \textbf{RQ1:} What is the benefit of integrating auxiliary data such as tweets, fact-checking URL's content and network structure?
\item \textbf{RQ2:} How helpful is adding SPPMI matrices of fact-checking URLs and guardians?
\item \textbf{RQ3:} What is the performance of the proposed $GAU$ model compared with other state-of-the-arts methods?
\item \textbf{RQ4:} What is the performance of the proposed GAU model for different types of guardians in terms of activeness level?
\item \textbf{RQ5:} What is the sensitivity of $GAU$ to hyperparameters?
\squishend

\subsection{Experimental Settings}

\noindent\textbf{Processing our dataset.}
We were interested in selecting active and professional guardians who frequently posted fact-checking URLs since they would be more likely to spread recommended fact-checking URLs than casual guardians.

Following a similar preprocessing approach to recommending scientific articles \cite{wang2011collaborative, wang2015collaborative}, we only selected guardians who used at least three distinct fact-checking URLs in their D-tweets and/or S-tweets. Altogether, 12,197 guardians were selected for training and evaluating recommendation models. They posted 4,834 distinct fact-checking URLs in total. The number of interactions was 68,684 (Sparsity:99.9\%). There were 9,710 D-guardians, 6,674 S-guardians and 4,187 users who played both roles. The total number of followers of the 12,197 guardians was 55,325,364, indicating their high impact on fact-checked information propagation. 


\smallskip
\noindent\textbf{Experimental design and metrics.} To validate our model, we followed a similar approach that \cite{liang2016factorization} did. In particular, we randomly selected 70\%, 10\% and 20\% URLs of each guardian for training, validation and testing. The validation data was used to tune hyper-parameters and to avoid overfitting. We repeated this evaluation scheme for five times, getting five different sets of training, validation and test data. The average results were reported. We used three standard ranking metrics such as Recall@k, MAP@k (Mean Average Precision) and NDCG@k (Normalized Discounted Cumulative Gain) \cite{manning2008introduction,liang2016factorization}. Since $k=10$ was used in \cite{abel2011analyzing}, we tested our model with $k\in \{5, 10, 15\}$.

\begin{table*}[]
	\centering
	\resizebox{1\linewidth}{!}
	{
		\begin{tabular}{l|lll|lll|lll|c}
			\hline
			Methods          & Recall@5    & NDCG@5      & MAP@5       & Recall@10   & NDCG@10     & MAP@10      & Recall@15   & NDCG@15     & MAP@15      & Avg. Rank \\ \hline
			BASIC         & 0.08919 \textcolor{blue}{(6)} & 0.06004 \textcolor{blue}{(6)} & 0.04839 \textcolor{blue}{(6)} & 0.13221 \textcolor{blue}{(6)} & 0.07417 \textcolor{blue}{(6)} & 0.05413 \textcolor{blue}{(6)} & 0.16208 \textcolor{blue}{(6)} & 0.08227 \textcolor{blue}{(6)} & 0.05653 \textcolor{blue}{(6)} & 6.0       \\ 
			BASIC+NW+UC            & 0.09967 \textcolor{blue}{(4)} & 0.06814 \textcolor{blue}{(5)} & 0.05535 \textcolor{blue}{(5)} & 0.14817 \textcolor{blue}{(4)} & 0.08399 \textcolor{blue}{(4)} & 0.06170 \textcolor{blue}{(5)} & 0.18280 \textcolor{blue}{(3)} & 0.09335 \textcolor{blue}{(5)} & 0.06432 \textcolor{blue}{(5)} & 4.4       \\
			BASIC+NW+UC+CSU        & 0.09900 \textcolor{blue}{(5)} & 0.06822 \textcolor{blue}{(4)} & 0.05604 \textcolor{blue}{(4)} & 0.14688 \textcolor{blue}{(5)} & 0.08386 \textcolor{blue}{(5)} & 0.06235 \textcolor{blue}{(4)} & 0.18266 \textcolor{blue}{(4)} & 0.09354 \textcolor{blue}{(4)} & 0.06522 \textcolor{blue}{(4)} & 4.3       \\
			BASIC+CSU+CSG          & 0.10247 \textcolor{blue}{(3)} & 0.06958 \textcolor{blue}{(3)} & 0.05670 \textcolor{blue}{(3)} & 0.14950 \textcolor{blue}{(3)} & 0.08497 \textcolor{blue}{(3)} & 0.06293 \textcolor{blue}{(3)} & 0.18205 \textcolor{blue}{(5)} & 0.09380 \textcolor{blue}{(3)} & 0.06554 \textcolor{blue}{(3)} & 3.2       \\
			BASIC+NW+UC+CSU+CSG    & 0.11133 \textcolor{blue}{(2)} & 0.07422 \textcolor{blue}{(2)} & 0.05978 \textcolor{blue}{(2)} & 0.16127 \textcolor{blue}{(2)} & 0.09065 \textcolor{blue}{(2)} & 0.06646 \textcolor{blue}{(2)} & 0.19516 \textcolor{blue}{(2)} & 0.09980 \textcolor{blue}{(2)} & 0.06917 \textcolor{blue}{(2)} & 2.0       \\
			Our GAU model & 0.11582 \textcolor{blue}{(1)} & 0.07913 \textcolor{blue}{(1)} & 0.06481 \textcolor{blue}{(1)} & 0.16400 \textcolor{blue}{(1)} & 0.09489 \textcolor{blue}{(1)} & 0.07118 \textcolor{blue}{(1)} & 0.19693 \textcolor{blue}{(1)} & 0.10381 \textcolor{blue}{(1)} & 0.07382 \textcolor{blue}{(1)} & 1.0       \\ \hline 
		\end{tabular}
	}
	\caption{Effectiveness of using auxiliary information and co-occurrence matrices. The GAU model outperforms the other variants significantly with p-value<0.001. }
	\label{tbl:model_evaluation}
	\vspace{-20pt}
\end{table*}

\subsection{Baselines and Our Model}
\label{sec:baselinse}
We compared our proposed model with the following four state-of-the-art collaborative filtering algorithms:
\squishlist
\item \textbf{BPRMF} Bayesian Personalized Ranking Matrix Factorization \cite{rendle2009bpr} optimizes the matrix factorization model with pairwise ranking loss. It is a common baseline for item recommendation.
\item \textbf{MF} Matrix Factorization (MF) \cite{koren2009matrix} is a standard technique in collaborative filtering. Given an interaction matrix $\textbf{X} \in \mathbb{R}^{N\times M}$, it factorizes \textbf{X} into two matrices $\textbf{U} \in \mathbb{R}^{N\times D}$ and $\textbf{V}\in \mathbb{R}^{D\times M}$, which are latent representations of users and items, respectively. 
\item \textbf{CoFactor} CoFactor \cite{liang2016factorization} extended Weighted Matrix Factorization (WMF) by jointly decomposing interaction matrix $\textbf{X}$ and co-occurrence SPPMI matrix for items (i.e., fact-checking URLs in this context). We set a confidence value $c_{X_{ij}=1}=1.0$ for $X_{ij}=1$, and we set $c_{X_{ij}=0}=0.01$ for non-observed interaction. The number of negative samples $s$ was grid-searched in a set $s\in\{1,2,5,10,50\}$, following the same settings as in \cite{liang2016factorization}. 
\item \textbf{CTR} Collaborative Filtering Regression \cite{wang2011collaborative} employed content of URLs (i.e., fact-checking pages in this context) to recommend scientific papers to users. Following exactly the best setting reported in the paper, we selected the top 8,000 words from fact-checking URLs' contents based on the mean of tf-idf values and set $\lambda_u=0.01$, $\lambda_v=100$, D=200, a=1 and b=0.01. 
\squishend

To build our GAU model, we conducted the grid-search to select the best value of $\alpha$, $\beta$ and $\gamma$ in $\{0.02, 0.04, 0.06, 0.08\}$. The number of negative samples $s$ for constructing SPPMI matrices was in $\{1,2,5,10,50\}$. For all of the baselines and the GAU model, we set latent dimensions to $D=100$ unless explicitly stated, and regularization value $\lambda$ was grid-searched in $\{10^{-5}, 3\times 10^{-5}, 5\times 10^{-5}, 7\times 10^{-5}\}$ by default. We only report the best result of each baseline.

We also attempted to compare our proposed model with content-based recommendation algorithms \cite{yamaguchi2013recommending, abel2011analyzing, abel2013twitter, abel2011semantic}. These methods mostly required collecting additional data from external data sources which are very time-consuming and expensive, and sometimes impossible for the third party researchers. We tried to compare our model with recent work \cite{yamaguchi2013recommending} and collected 5,383,598 followees of the 12,196 guardians and over 15 million distinct Twitter lists in which at least one of the followees was included. However, we were not able to collect all fact-checking tweets posted by these followees during the same data collection period (from May 16, 2016 to July 7, 2017). Therefore, we only used followees that were in the set of 12,197 guardians. But, maybe because of the limited data, it performed poorly in the experiments. Therefore, we omit its results in the experiments. Instead, we report performance of our GAU model and the four state-of-the-art collaborative filtering algorithms.
\subsection{Effectiveness of Auxiliary Information and SPPMI Matrices (RQ1 \& RQ2)}
Before comparing our GAU model with the four baselines, we first examined the effectiveness of exploiting auxiliary information and the utility of jointly factorizing SPPMI matrices. Starting from our basic model in Eq.\ref{eq:basic}, we created variants of the $GAU$ model. Since there are many variants of $GAU$, we selectively report performance of the following $GAU$'s variants:
\squishlist
\item Our basic model (Equation \ref{eq:basic}) (BASIC)
\item BASIC + Network + URL's content (BASIC+NW+UC)
\item BASIC + Network + URL's content + URL's SPPMI matrix (BASIC+NW+UC+CSU)
\item BASIC + URL's SPPMI matrix + Guardians' SPPMI matrix (BASIC+CSU+CSG)
\item BASIC + Network + URL's content + SPPMI matrix of URLs + SPPMI matrix of Guardians (BASIC+NW+UC+CSU+CSG)
\item Our GAU model 
\squishend
\begin{figure*}[h]
	\centering
	\subfigure[Recall@k]{
		\label{fig:recall_3_left_models}
		\includegraphics[trim=1.5cm 1cm 6cm 2cm,clip,width=0.3\linewidth, height=1.1in]{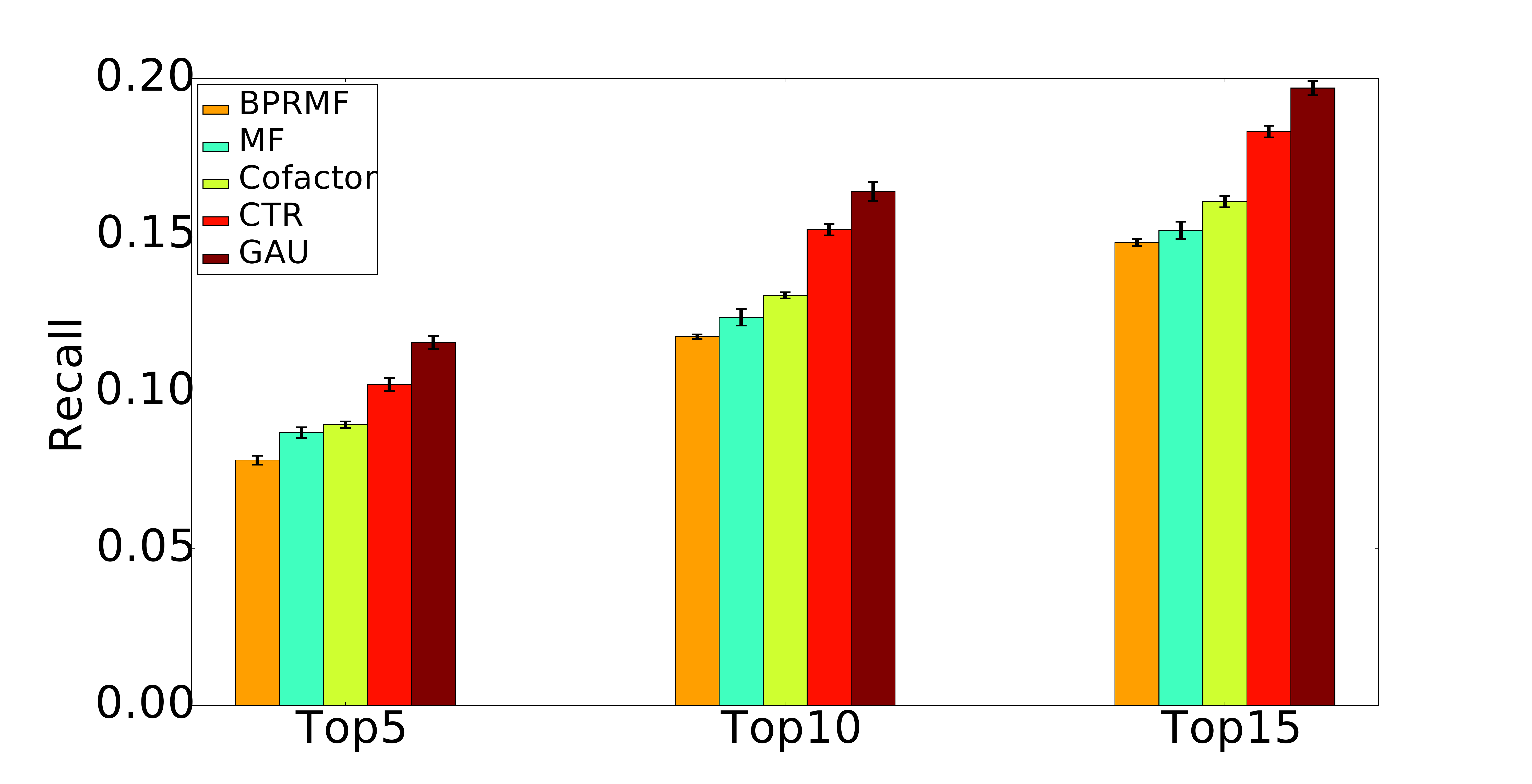}
	}
	\subfigure[NDCG@k]{
		\label{fig:ndcg_3_left_models}
		\includegraphics[trim=1.5cm 1cm 6cm 2cm,clip,width=0.3\linewidth, height=1.1in]{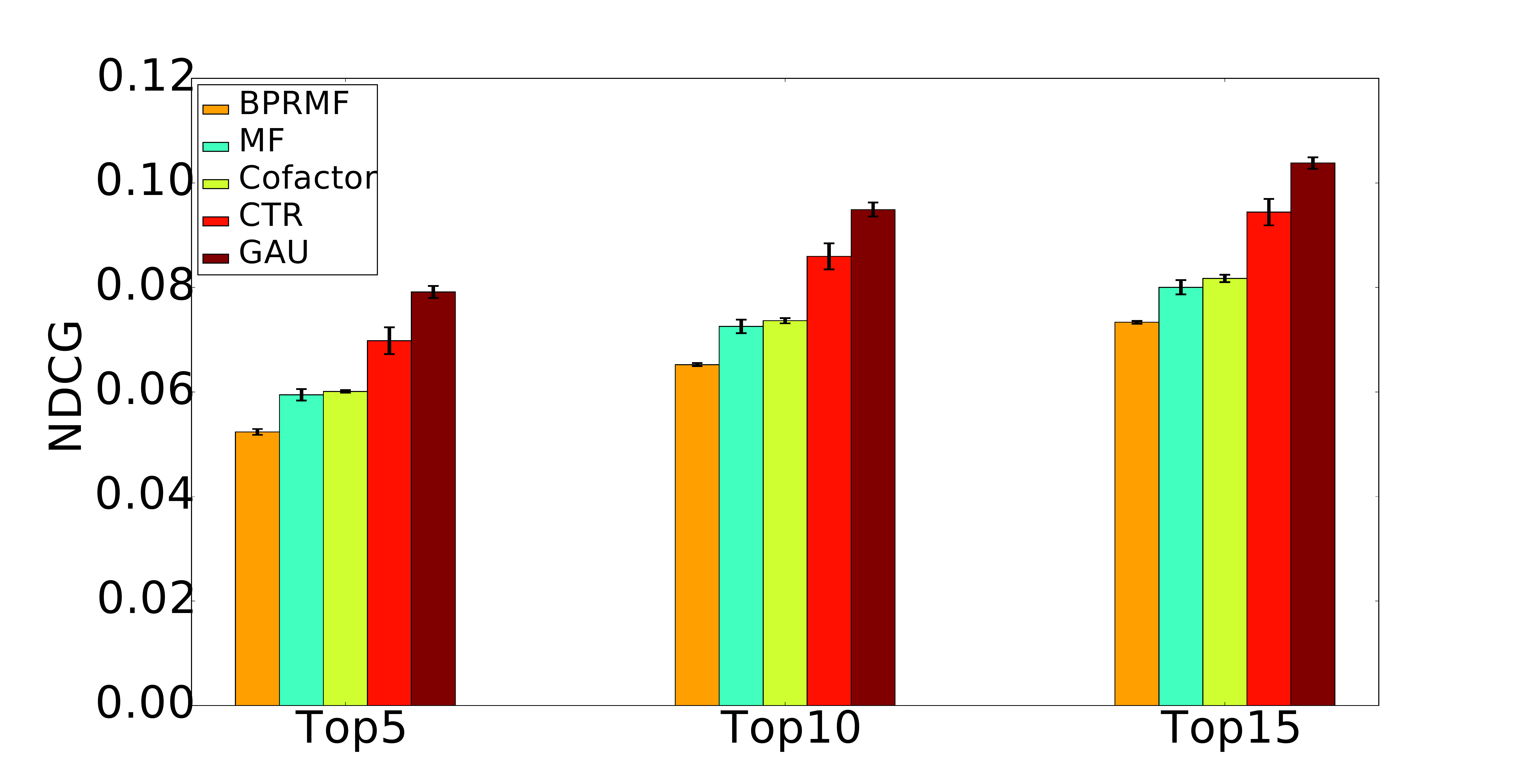}
	}
	\subfigure[MAP@k]{
		\label{fig:map_3_left_models}
		\includegraphics[trim=1.5cm 1cm 6cm 2cm,clip,width=0.3\linewidth, height=1.1in]{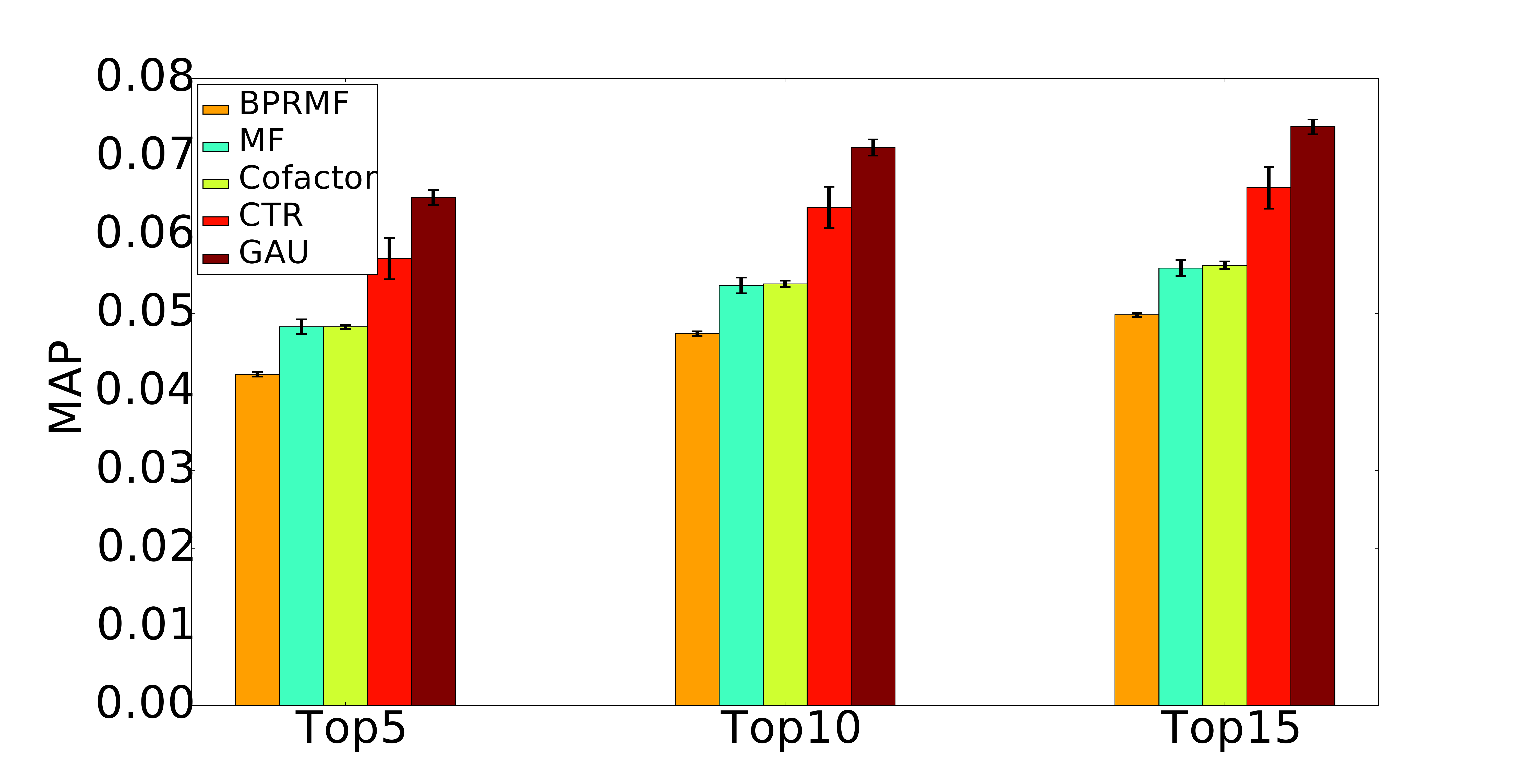}
	}
	\vspace{-10pt}
	\caption[]{Performance of our GAU model and 4 baselines. The GAU model outperforms the baselines (p-value<0.001).}
	\label{fig:last_comparisons}
	\vspace{-10pt}
\end{figure*}

\begin{figure*}[t]
	\centering
	\subfigure[Recall@15]{
		\label{fig:recall@15_cold_warm_hot}
		\includegraphics[trim=1.5cm 1cm 6cm 2cm,clip,width=0.3\linewidth, height=1.1in]{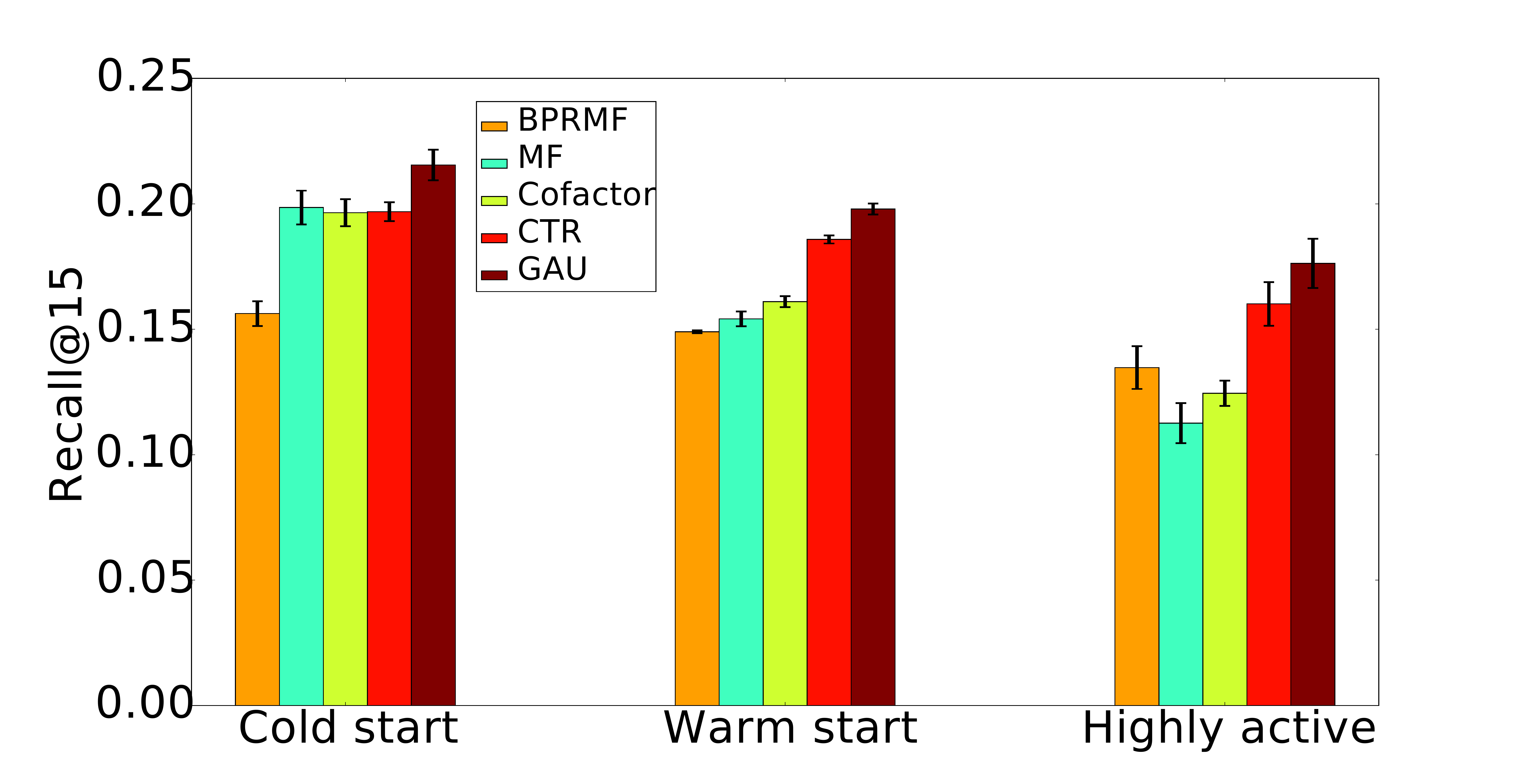}
	}
\subfigure[NDCG@15]{
	\label{fig:ndcg@15_cold_warm_hot}
	\includegraphics[trim=1.5cm 1cm 6cm 2cm,clip,width=0.3\linewidth, height=1.1in]{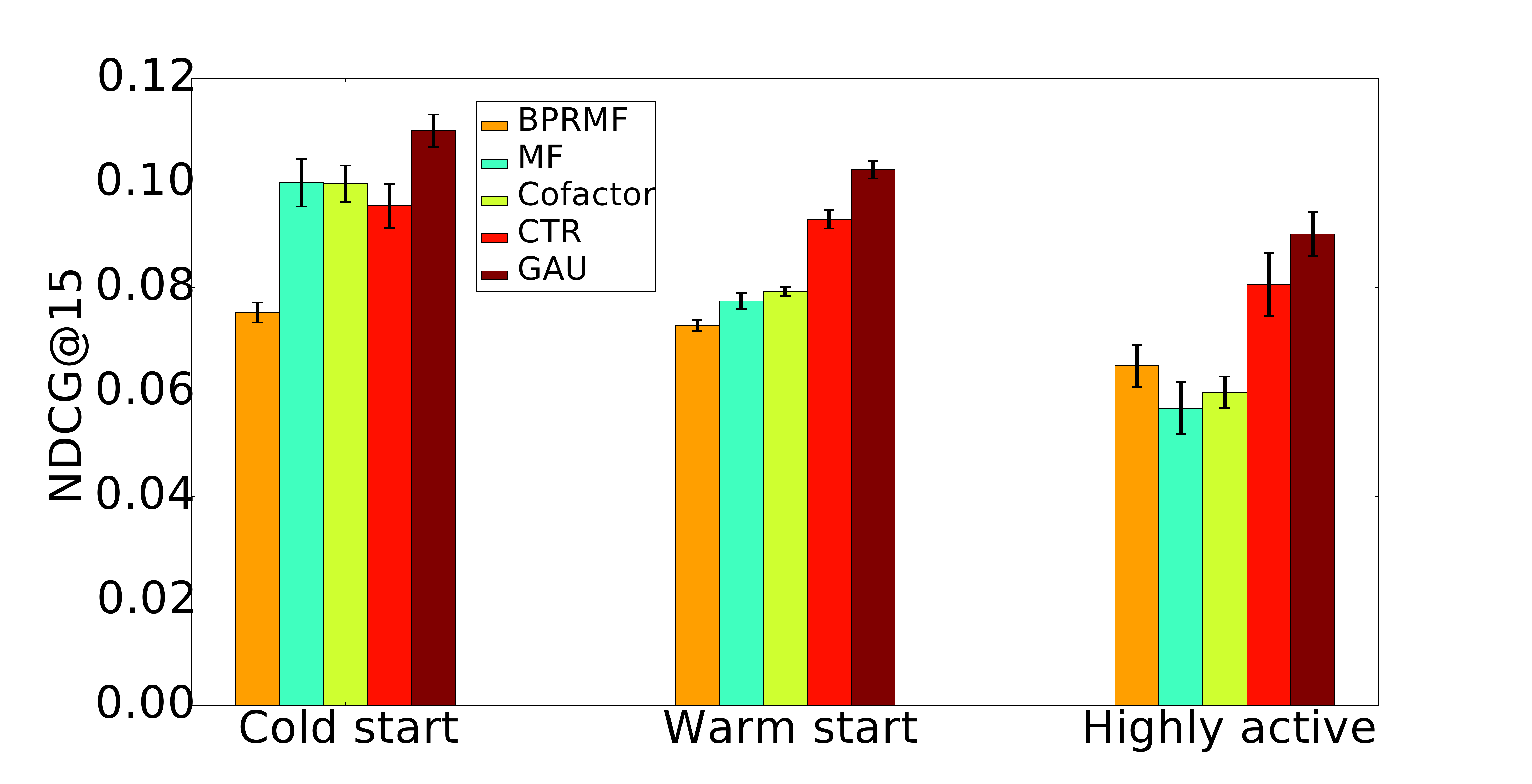}
}
\subfigure[MAP@15]{
	\label{fig:map@15_cold_warm_hot}
	\includegraphics[trim=1.5cm 1cm 6cm 2cm,clip,width=0.3\linewidth, height=1.1in]{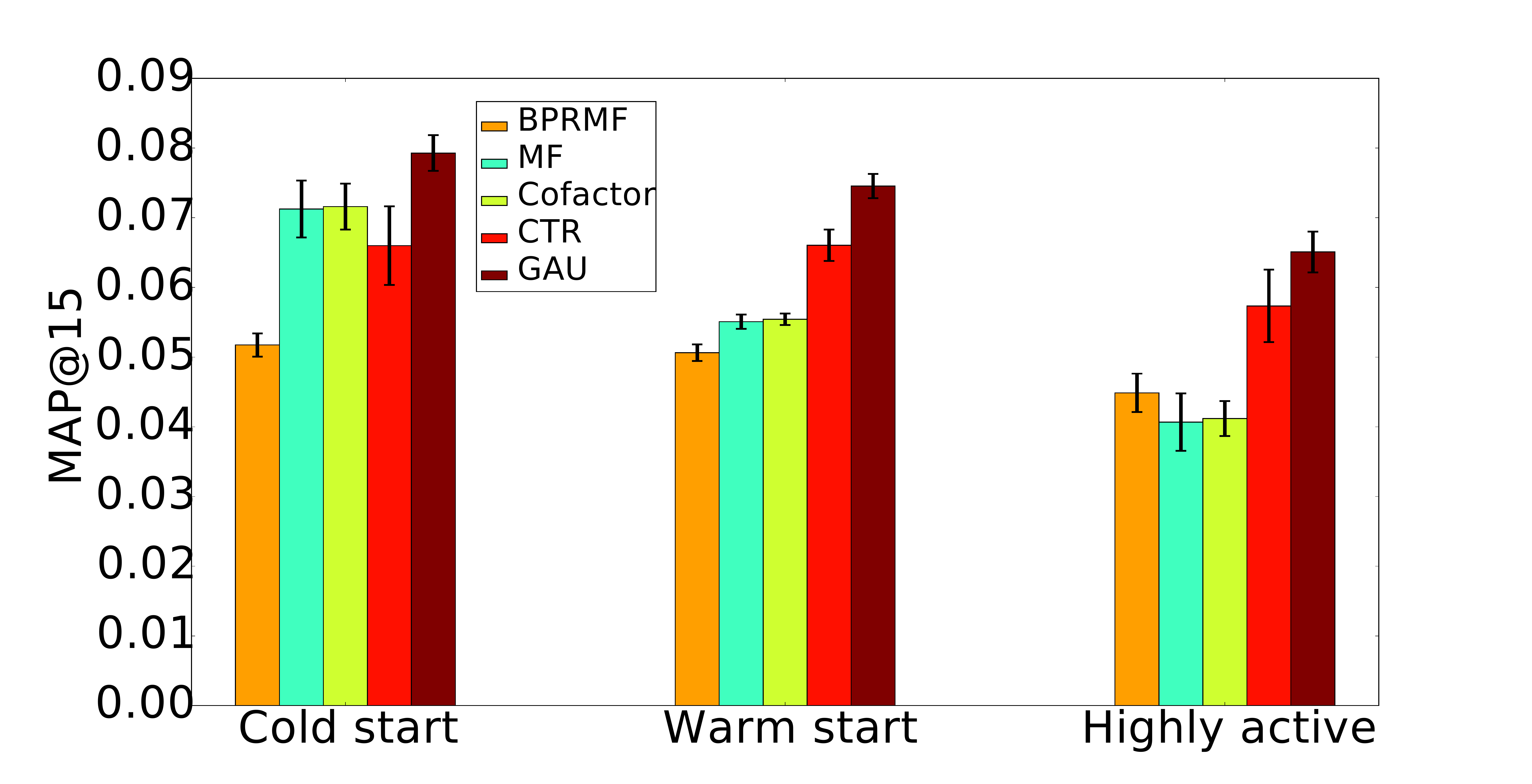}
}
	\vspace{-11pt}
	\caption[]{Performance of GAU and baselines for three types of guardians. GAU outperforms the baselines (p-value<0.01).}
	\label{fig:cold_warm_hot_start_guardians}
	\vspace{-11pt}
\end{figure*}

Table \ref{tbl:model_evaluation} shows performance of the variants and the GAU model. It shows the rank of each method based on reported metrics. 
 By adding social network information and fact-checking URL's content to Equation \ref{eq:basic}, there was a huge climb in performance of BASIC+NW+UC over BASIC across all metrics. In particular, Recall, NDCG and MAP of BASIC+NW+UC were better than BASIC about 12.20\%$\pm$1.31\%, 13.39\%$\pm$0.34\% and 14.04\%$\pm$0.76\%, respectively (confidence interval 95\%). 
 These results confirm the effectiveness of exploiting auxiliary information. 

 How about using co-occurrence SPPMI matrices of fact-checking URLs and guardians? First, when adding co-occurrence SPPMI matrix of fact-checking URL (CSU) to the variant BASIC+NW+UC, we did not see much improvement across all settings. Second, when jointly factorizing two SPPMI matrices (BASIC+CSU+CSG) and comparing it with the variant BASIC+NW+UC, we can see that BASIC+CSU+CSG and BASIC+NW+UC performed equally well. Again, BASIC+CSU+CSG did not use any additional data sources except the interaction matrix \textbf{X}. It is an attractive benefit since it did not depend on other data sources. In other words, it reflects that regularizing the BASIC model with SPPMI matrices is comparable to adding network data and URLs' contents to the BASIC model.

 So far, both auxiliary information and SPPMI matrices are beneficial to improving recommendation quality. How about combining all of them into a single model? Will performance be further improved? We turned to the variant BASIC+NW+UC+CSU+CSG. As expected, BASIC+NW+UC+CSU+CSG enhanced CSU+CSG by 7.90\%$\pm$1.79\% Recall, 6.58\%$\pm$0.40\% NDCG, and 5.53\%$\pm$0.22\% MAP. Its results were also higher than BASIC+NW+UC about 9.10\%$\pm$6.15\% Recall, 7.92\%$\pm$2.50\% NDCG and 7.75\%$\pm$0.58\% MAP. 

Since adding auxiliary data was valuable, we now exploit another data source -- 200 recent tweets' content. Consistently, adding the tweets' content indeed improved performance. The improvement of the GAU over BASIC+NW+UC+CSU+CSG model was $4.0\%$ Recall, $6.6\%$ NDCG and $8.4\%$ MAP. This improvement is statistically significant with p-value<0.001 using Wilcoxon one-sided test. Comparing the GAU with the BASIC model, we observed a dramatic increase in performance across all metrics. Specifically, Recall, NDCG and MAP were improved by 25.13\%$\pm$10.64\%, 28.64\%$\pm$7.13\% and 32\%$\pm$4.29\% respectively.

Based on the experiments, we conclude that auxiliary data as well as co-occurrence matrices are helpful to improve recommendation quality. Adding CSU+CSG or NW+UC enhanced the BASIC model by 12\% to 14\%. Our GAU model performed best, which improved the BASIC model by 25\%$\sim$32\%.

\subsection{Performance of GAU and Baselines (RQ3)}
Figure \ref{fig:last_comparisons} shows the performance of the four baselines and GAU. MF was better than BPRMF which was designed to optimize Area Under Curve (AUC). Similar results were reported in \cite{wu2016collaborative}. CTR was a very competitive baseline. This reflects the importance of fact-checking URL's content (i.e., fact-checking page) in recommending right fact-checking URLs to guardians. GAU performed better than CTR by $12.75\%\pm 0.95\%$ Recall, $11.2\%\pm 4.6\%$ NDCG, and $12.5\%\pm 2.5\%$ MAP. GAU also outperformed CoFactor with a large margin by $25.8\%\pm 8.4\%$ Recall, $29.2\%\pm 5.8\%$ NDCG, and $32.6\%\pm 3.4\%$ MAP (confidence interval 95\%). Overall, our GAU model significantly outperformed all the baselines (p-value<0.001). The improvement over the baselines was 11\%$\sim$33\%.

\subsection{Performance of Models for Different Types of Guardians (RQ4)}
We grouped guardians into three types based on the number of their fact-checking URLs (i.e., the activeness level) to see whether our GAU still outperforms the baselines in all the three types. By sorting guardians in the ascending order of the number of their fact-checking URLs, we annotated the first 20\% guardians as cold-start guardians, the next 60\% guardians as warm-start guardians, and the last 20\% guardians as highly active guardians.

Figure \ref{fig:cold_warm_hot_start_guardians} shows performance of GAU and the baselines in Top 15 results. A general pattern of all the models was that they performed pretty well for cold-start guardians, and their performance slightly decreased as guardians posted more fact-checking URLs. We observed consistent results in top 5 and top10 as well.


GAU outperformed CTR in cold-start, warm-start and highly active guardians, improving Recall@15 by 6.5\%$\sim$10.0\%, NDCG@15 by 10.2\%$\sim$15.0\%, and MAP@15 by 12.8\%$\sim$20.1\%. Overall, GAU consistently outperformed the baselines for all three groups according to the three metrics. Its improvement was about 6.5\%$\sim$20.1\%.

\subsection{Exploiting hyper-parameters (RQ5)}
We investigated the impact of hyper-parameters $\alpha$, $\beta$ and $\gamma$ on the GAU model. These hyper-parameters control the contribution of social network, fact-checking URL's content and 200 recent tweets' content to the GAU. We tested $\alpha$, $\beta$ and $\gamma$ from 0.01 to 0.09, increasing 0.01 in each step, and then report the average recall@15, while we fixed $\lambda=3\times 10^{-5}$ and the number of negative samples $s=10$.

\begin{figure}[t]
	\centering
	\subfigure[$\beta=0.08$]{
		\label{fig:fixed_beta}
		\includegraphics[trim=3cm 7cm 3cm 8cm,clip,width=1.0in, height=0.9in]{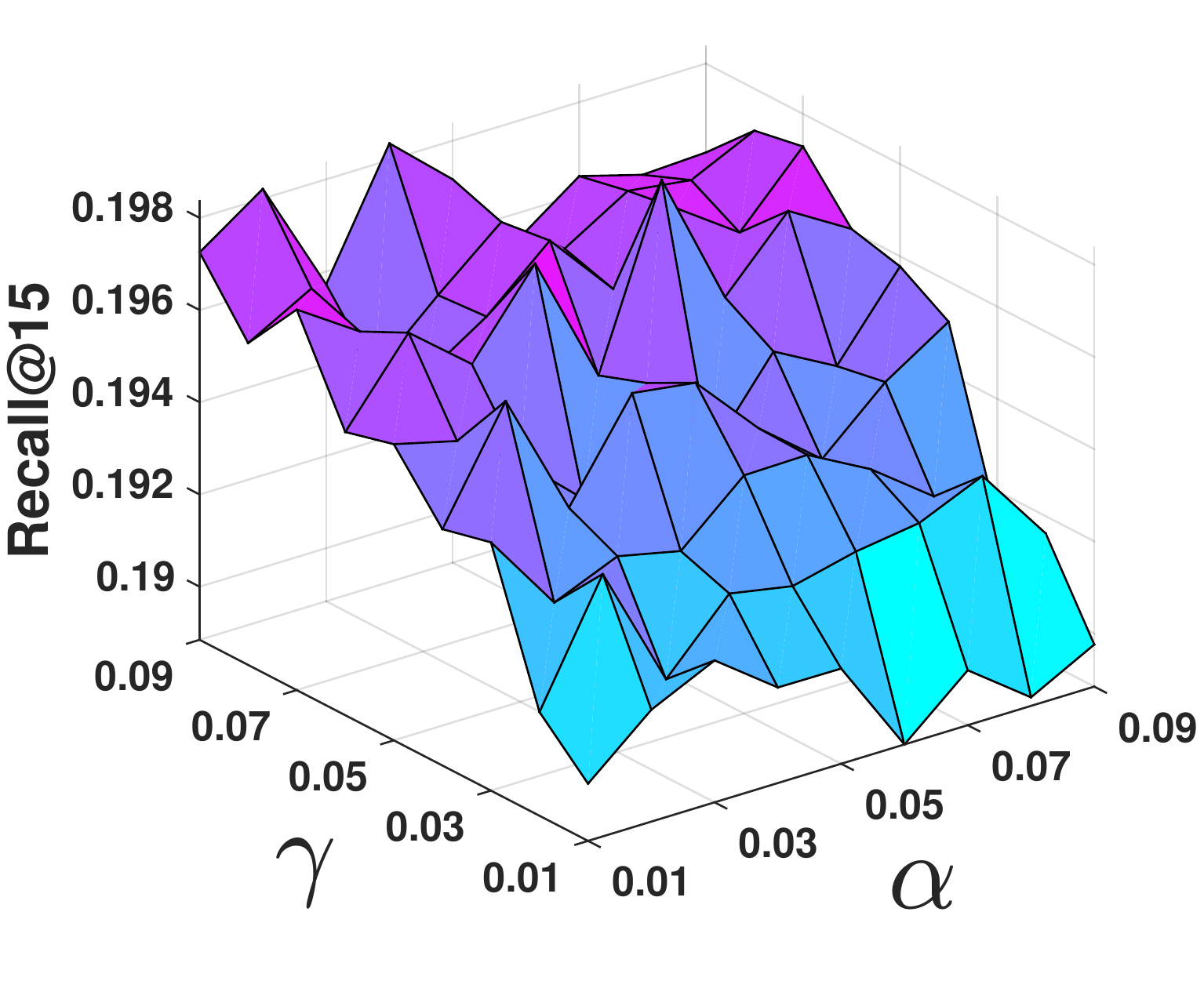}
	}
	\subfigure[$\alpha=0.08$]{
		\label{fig:fixed_alpha}
		\includegraphics[trim=3cm 7cm 3cm 8cm,clip,width=1.0in, height=0.9in]{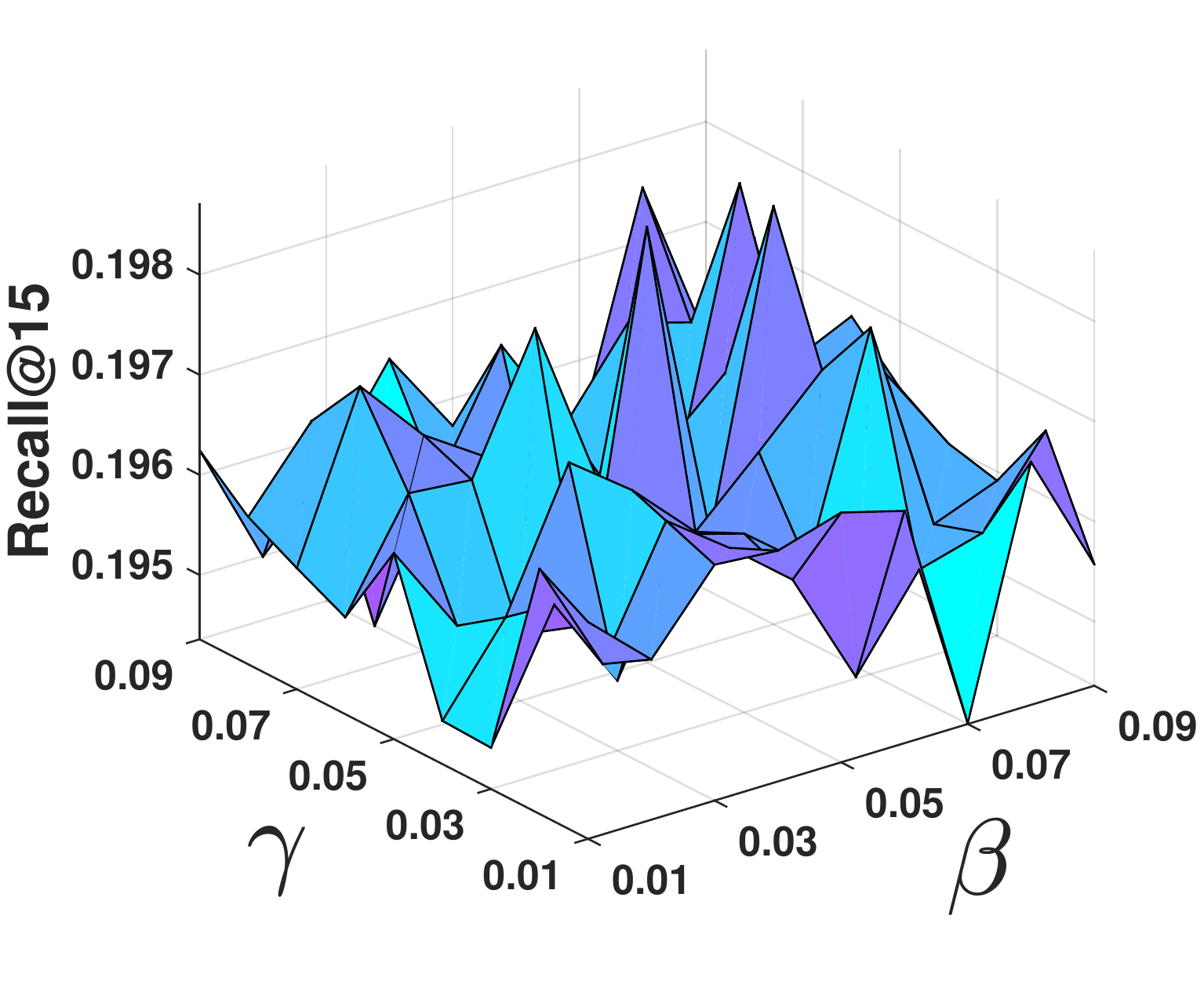}
	}
	\subfigure[$\gamma=0.08$]{
		\label{fig:fixed_gamma}
		\includegraphics[trim=3cm 7cm 3cm 8cm,clip,width=1.0in, height=0.9in]{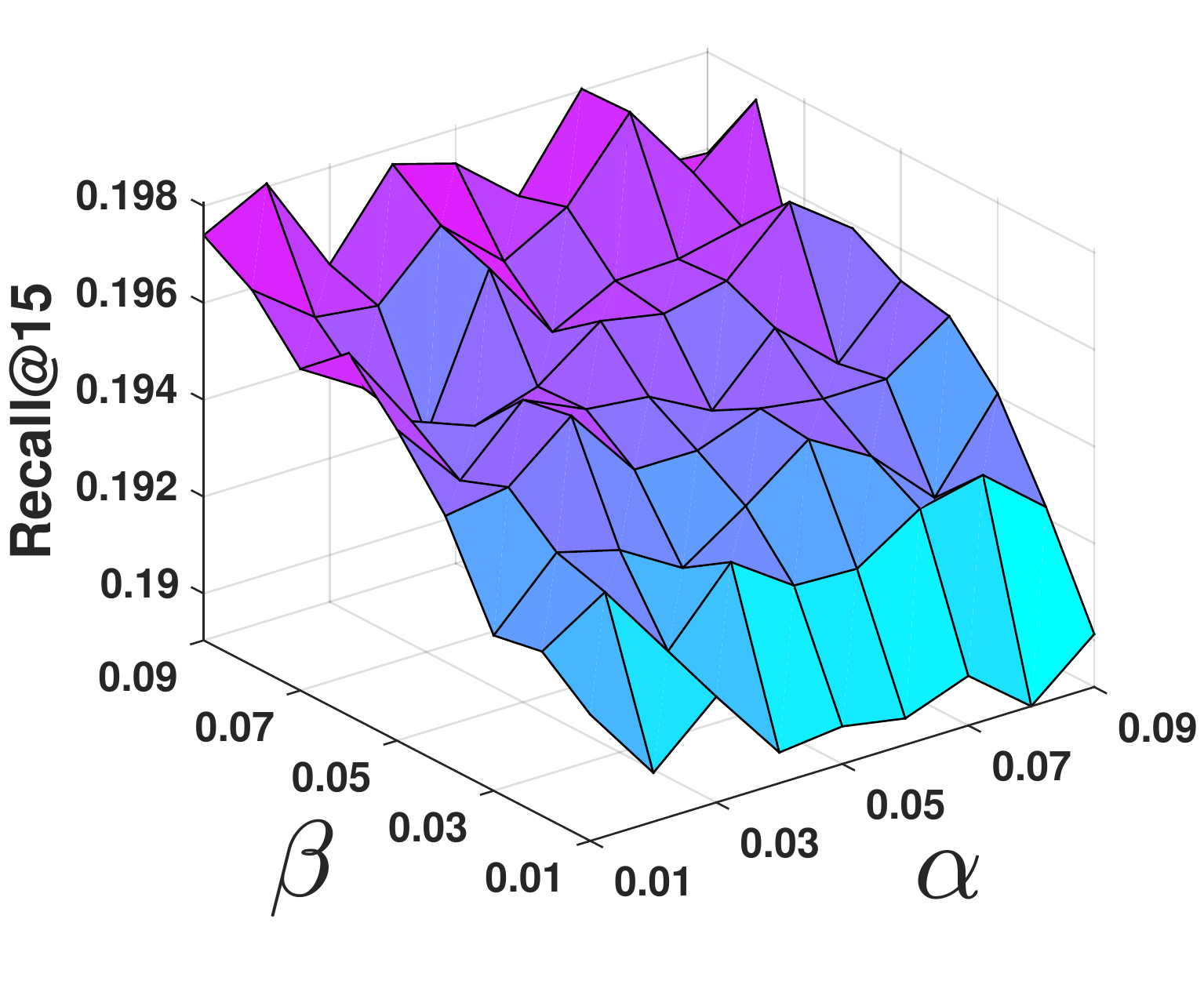}
	}
	\vspace{-10pt}
	\caption[]{Hyper-parameter sensitivity.}
	\label{fig:3d_plots}
	\vspace{-15pt}
\end{figure}

In Figure \ref{fig:fixed_beta}, we fixed $\beta=0.08$ and varied $\alpha$ and $\gamma$. The general trend was that recall@15 gradually went up, when $\alpha$ and $\gamma$ increased. It reached the peak, when $\alpha=0.06$ and $\gamma=0.06$. Next, we fixed $\alpha=0.08$. It seems recall@15 fluctuated when varying $\beta$ and $\gamma$, but the amplitude was small. The max Recall@15 was only 2.2\% larger than the smallest Recall@15. Finally, $\gamma$ was fixed to 0.08. The trend was similar to Figure \ref{fig:fixed_beta}. In general, when $\alpha, \beta$ and $\gamma$ are large, the performance tends to improve, which suggests the importance of regularizing our model using the auxiliary information.

\section{Discussion}
In Section \ref{sec:data_analysis}, we showed that guardians had great enthusiasm for information credibility. Nevertheless, many guardians only posted 1$\sim$2 fact-checking tweets. Therefore, we only recommended URLs to highly enthusiastic guardians, who posted at least 3 fact-checking URLs, because they may continue to be active in spreading fact-checked information in the future. Another observation is that the top verified guardians seem not to be active in the covered time period. We conjecture that these verified guardians may be cautious about what they should post to their followers \cite{marwick2011see}. We also showed that exploiting auxiliary information indeed helped improve recommendation quality. There is considerable potential to integrate other data sources such as temporal factors and activeness of guardians to further improve the proposed recommender system. We leave them for future work.

\section{Conclusion}
We collected a list of guardians, who showed their interests in information credibility by embedding fact-checking URLs in their posts. The guardians were very active in posting credible information and were mostly interested in politics, fauxotography and fake news. After analyzing our dataset, we proposed a recommendation model to personalize fact-checking URLs to the guardians toward enhancing their engagement in fact-checking activities and encouraging them to post more credible information. Our proposed model outperformed four baselines (i.e., MF, CoFactor, BPRMF and CTR). In future work, we will upgrade our model to address the cold-start issue where guardians posted less than 3 fact-checking URLs and will investigate whether employing deep learning techniques would further improve performance of our model.
\section*{Acknowledgment}
This work was supported in part by NSF grant CNS-1755536, Google Faculty Research Award, and Microsoft Azure Research Award. Any opinions, findings and conclusions or recommendations expressed in this material are the author(s) and do not necessarily reflect those of the sponsors.



%
%

\bibliographystyle{ACM-Reference-Format}
\bibliography{www}

\end{document}